\documentclass[12pt,preprint]{aastex}

\shorttitle{Cosmic-Ray Background Flux Model}
\shortauthors{T. Mizuno et al.}

\begin{document}


\title{Cosmic-Ray Background Flux Model based on a
Gamma-Ray Large-Area Space Telescope 
Balloon Flight Engineering Model}


\author{T. Mizuno, T. Kamae, G. Godfrey and T. Handa\altaffilmark{1}}
\affil{Stanford Linear Accelerator Center, 2575 Sand Hill Road,
Menlo Park, CA 94025, USA}
\email{mizuno@SLAC.Stanford.EDU}

\author{D. J. Thompson}
\affil{NASA Goddard Space Flight Center, Greenbelt, MD 20771, USA}

\author{D. Lauben}
\affil{Hansen Experimental Physics Laboratory, 
Stanford University, Stanford, CA 94305, USA}

\author{Y. Fukazawa}
\affil{Department of Physics, Hiroshima University, 1-3-1 Kagamiyama,
Higashi-Hiroshima, Hiroshima 739-8526, Japan}

\and

\author{M. Ozaki}
\affil{Institute of Space and Astronautical Science,
Japan Aerospace Exploration Agency,
1-3-1 Yunodai, Sagamihara, Kanagawa 229-8510, Japan}


\altaffiltext{1}{
present address: 
Computer Software Development Co., Ltd.,
2-16-5 Konan, Minato-ku,
Tokyo 108-0075, Japan}


\begin{abstract}
Cosmic-ray background fluxes were modeled based on existing measurements
and theories and are presented here.
The model, originally developed for the 
Gamma-ray Large Area Space Telescope (GLAST) Balloon Experiment,
covers the entire solid angle (${\rm 4\pi~sr}$),
the sensitive energy range of the instrument
(${\rm \sim 10~MeV~to~100~GeV}$) and abundant components
(proton, alpha, $e^{-}$, $e^{+}$, $\mu^{-}$, $\mu^{+}$ and gamma).
It is expressed in analytic functions in which
modulations due to the solar activity and the Earth geomagnetism are
parameterized.
Although the model is intended to be used primarily 
for the GLAST Balloon Experiment, 
model functions in low-Earth orbit are also
presented and can be used for other high energy
astrophysical missions.
The model has been validated via comparison with the data of
the GLAST Balloon Experiment.
\end{abstract}



\keywords{gamma rays --- instruments --- balloons --- cosmic rays}


\section{Introduction}

In high energy gamma-ray astrophysics observations,
it is vital to reduce the background effectively in order 
to achieve a high sensitivity,
for the source intensity is quite low.
Most of the background events are generated via an interaction between
cosmic-rays and the instrument, and an accurate
cosmic-ray flux model and computer simulation are necessary
to develop the background rejection algorithm and 
evaluate the remaining background level.
The model needs to cover the entire solid angle (${\rm 4 \pi~sr}$)
and the sensitive energy range of the instrument.
While there are many measurements on cosmic-rays
reported in literature,
they are fragmentary and need to be compiled, reviewed and validated.

The Large Area Telescope (LAT) is the high-energy gamma-ray detector
onboard the Gamma-ray Large Area Space Telescope (GLAST) satellite
to be launched to low-Earth orbit 
(altitude of ${\rm \sim 550~km}$) in 2007 
\citep{GLAST}.
It consists of 16 towers in a 4 x 4 array.  
Each tower has a set of high-Z foil and Si strip
pair conversion trackers (TKR) and a CsI(Tl) hodoscopic calorimeter (CAL). 
The towers are surrounded on the top and sides 
by a set of plastic scintillator
anticoincidence detectors (ACD).
Because the LAT will have a large active volume
(${\rm 1.8~m \times 1.8~m \times 1~m}$),
the data acquisition trigger, based on signals in three consecutive
x-y pair layers of the TKR, is expected to be about 10~kHz,
whereas the gamma-ray signal from a strong extraterrestrial source
(e.g., Crab pulsar) is only about one per minute.
In order to achieve the required sensitivity, we have to develop
a ``cosmic-ray proof'' instrument
by an optimized on-line/off-line
event filtering scheme.
All possible background types need to be studied in detail
by beam tests and computer simulations.

To validate the basic design of LAT in a space-like
environment, and collect particle incidences to be used
as a background event database for LAT,
a Balloon Flight Engineering Model (BFEM),
representing one of 16 towers that compose the LAT, 
was built and launched on
August 4, 2001 \citep{Thompson2002}.
The instrument operated successfully and took data for about
two hours during the ascent and three hours during the level flight.
A Geant4-based \citep{Geant4} simulator of 
BFEM was constructed for this balloon experiment.
We also have developed a cosmic-ray background flux model
based on existing measurements and theoretical predictions,
and validated the model by comparing with the data of BFEM.
The model covers the entire solid angle
($\rm 4 \pi~sr$), a wide energy range between 10~MeV to 100~GeV
and abundant components 
(proton, alpha, $e^{\pm}$, $\mu^{\pm}$ and gamma ray).
In this paper, we present the model functions as 
``a working model of cosmic-ray background fluxes'' for 
high energy astrophysical missions.
Although the models are intended to be used primarily for GLAST BFEM,
they are expressed in analytic functions in which
modulations due to solar activity and Earth geomagnetism are 
parameterized, and they are easy to be applied to other
balloon and satellite experiments.
The models in low-Earth orbit are being used to simulate the
GLAST LAT in the orbital environment and to develop
background rejection algorithms.

\section{Instrument and Simulator}

The GLAST LAT BFEM was a reconfigured version of 
the Beam Test Engineering Model 
(BTEM) that represents one of 16 towers of the LAT and
consists of ACD, TKR and CAL.
The BFEM differed from BTEM in that 6 silicon tracker layers
were removed and the data acquisition system was reconfigured.
The BTEM was tested with $e^{-}$, $e^{+}$ and gamma beams
at the Stanford Linear Accelerator Center (SLAC), and the
results were
published (\citet{BTEM}).
The BFEM ACD consisted of 13 segmented plastic anticoincidence
scintillators
to help identify the background events due to charged particles.
Among them, 4 tiles were placed above the TKR and 8
on the side. In addition, one big tile was placed
above 4 top tiles to cover gaps between tiles.
The TKR consisted of 13 x-y layers of Si strip detectors (SSDs)
of $400\mu{\rm m}$ thickness
to measure particle tracks and 11 lead foils
to convert gamma-rays.
The area of SSDs was $\rm 32\ cm \times 32\ cm$ for each layer,
but the top 5 x-y pairs had smaller area.
Among 11 lead foils, the top 8 were 
3.6\% radiation length thick, the following three were
28\% radiation length thick,
and the last two x-y layers had no lead converter.
The instrument trigger was provided
by hits in three x-y pairs in-a-row (six-fold coincidence).
The CAL consisted of 80 CsI crystals 
hodoscopically laid out 
(eight layers of 10 logs in x or y directions)
and measured energy deposit and shower profile.
A set of four plastic scintillators called
eXternal Gamma-ray Target (XGT) was mounted above
the instrument to tag 
cosmic-ray interactions in the scintillators and
constrain the gamma-ray origin.
The four detector components,
readout electronics and support structures were
mounted in a Pressure Vessel (PV) because
not all of the electronics and data recorders 
were designed to be operated in vacuum.

We developed a Monte-Carlo simulation program of the BFEM
with a Geant4 toolkit. The geometry of the BFEM
implemented in the simulator is shown in Figure~1:
support structures and the pressure vessel as well as detectors
were implemented.
Standard electromagnetic processes, hadronic interactions
and decay processes were simulated.
Cutoff length 
(range threshold of secondary particles to be generated) was set at 
$10\mu m$ in TKR and 1~mm for other regions.
We utilized Geant4 ver 5.1 with patch 01.

\section{Cosmic-Ray Background Flux Models}
Cosmic-rays in or near the Earth atmosphere
consist of the primary and the secondary components.
The primary cosmic-rays are generated in and propagate 
through extraterrestrial space. 
The main component is known to be protons.
When primary particles penetrate into the air and interact with
molecules, they produce relatively low-energy particles,
i.e., secondary cosmic-rays.
The first step in building a working model of cosmic-ray fluxes
is to represent the existing measurements and theoretical predictions
by analytic functions. Below we describe the model functions
particle by particle.
Data are very scarce or absent for large zenith angle and low energy,
and expected uncertainties are also mentioned.

\subsection{Primary Charged Particle Spectra}
Primary particles that reach the top of the atmosphere are believed to come
predominantly from outside the solar system.
Their spectra are known to be affected by the solar activity 
and Earth geomagnetic field.
The spectra of the primaries in interstellar space
can be modeled by a power-law in rigidity (or momentum) as
\begin{equation}
Unmod(E_{\rm k}) = 
A \times \left( \frac{R(E_{\rm k})}{\rm GV} \right)^{-a} 
\quad,
\end{equation}
where $E_{\rm k}$ and $R$ are the kinetic energy and rigidity
of the particle, respectively.
These particles are decelerated
by solar wind as they enter the solar system and hence
their flux shows anti correlation with the solar activity.
To model this solar modulation, we adopted
the formula given by \citet{Gleeson1968}:
the modulation is
expressed by an effective shift of energy as
\begin{equation}
Mod(E_{\rm k}) = 
Unmod (E_{\rm k}+Ze\phi)
\times \frac{\left( E_{\rm k} + Mc^{2} \right)^2 
- \left( Mc^{2}\right)^2}
{\left( E_{\rm k} + Mc^{2} + Ze\phi\right)^{2} 
- \left( Mc^{2}\right)^{2}}
\quad,
\end{equation}
where $e$ is the magnitude of electron charge,
$Z$ the atomic number of particle,
$M$ the particle mass and $c$ the speed of light.
A parameter $\phi$ is introduced to represent the solar modulation
which varies from $\sim 550~{\rm MV}$
for solar activity minimum to $\sim 1100~{\rm MV}$
for solar activity maximum.
Since the GLAST Balloon experiment took place in 2001 August,
we fixed $\phi$ at 1100~MV, a typical value for 
solar activity maximum.

The second and much bigger modulation is 
introduced as the
low energy cutoff due to the Earth geomagnetism.
We modeled this effect based on the AMS measurements
\citep{Alcaraz2000Proton} in which 
the proton flux was measured over a wide geomagnetic
latitude range. We defined the 
reduction factor as
\begin{equation}
\frac{1}{1+\left( \frac{R}{R_{\rm cut}} \right)^{-12.0}}
\quad.
\end{equation}
Here, $R_{\rm cut}$ is the cutoff rigidity 
calculated by assuming
the Earth magnetic field to be of dipole shape, as
\begin{equation}
R_{\rm cut} = 
14.9 \times \left(1+\frac{h}{R_{\rm Earth}} \right)^{-2.0}
\times \left( \cos\theta_{\rm M} \right)^{4}
\ {\rm GV}
\quad,
\end{equation}
where $h$ is the altitude, $R_{\rm Earth}$ the Earth radius
and $\theta_{\rm M}$ the geomagnetic latitude
\citep{Zombeck,HEAP}.
For the GLAST Balloon Experiment
they were $h\sim38$~km, $\theta_{\rm M} \sim 0.735$~rad, 
and $R_{\rm cut}=4.46~{\rm GV}$.
We also referred to $e^{-}$ measurements by
AMS \citep{Alcaraz2000Lepton} and found that electrons
are less affected by Earth geomagnetism.
Although the geomagnetic cutoff is expected to be the
same regardless of the particle type,
we gave priority to
the experimental data and used
\begin{equation}
\frac{1}{1+\left( \frac{R}{R_{\rm cut}} \right)^{-6.0}}
\quad,
\end{equation}
instead of equation~(3) for $e^{-}$/$e^{+}$.
The reduction factor of equations~(3) and (5) reproduces
the cutoff seen in AMS data in $0 \le \theta_{\rm M} \le 0.8~{\rm rad}$,
but gives lower flux in high
geomagnetic latitude region.

The primary spectrum is then expressed as
\begin{equation}
Primary(E_{\rm k}) = 
Unmod(E_{\rm k} + Ze\phi)
\times \frac{\left( E_{\rm k} + Mc^{2} \right)^2 
- \left( Mc^{2}\right)^2}
{\left( E_{\rm k} + Mc^{2} + Ze\phi\right)^{2} 
- \left( Mc^{2}\right)^{2}}
\times
\frac{1}{1+\left( \frac{R}{R_{\rm cut}} \right)^{-r}}
\quad,
\end{equation}
where $r=12.0$ and 6.0 for proton/alpha and $e^{-}$/$e^{+}$,
respectively.
This model function can be 
used for the entire solar cycle
and entire low-Earth orbit by adjusting $\phi$ and $R_{\rm cut}$.
We assume the angular distribution is uniform for
$0^{\circ} \le \theta \le \theta_{\rm c}$ and to be 0
for $\theta_{\rm c} \le \theta \le 180^{\circ}$,
where $\theta$ is the zenith angle from vertical.
The cutoff angle $\theta_{\rm c}$ is introduced
to represent the Earth horizon and is
$113^{\circ}$ at low-Earth orbit ($\sim 550~{\rm km}$).
The primary flux at balloon altitude is attenuated by air and the
effect is described in the following sections particle by particle.
Note that the east-west effect is not taken into account for simplicity:
we aim to construct models which provide the correct flux
integrated over the azimuth angle.

\subsection{Proton Flux}

Recent measurements by BESS \citep{BESS}
and AMS \citep{Alcaraz2000Proton, Alcaraz2000Proton2}
provided us with an accurate spectrum of primary protons.
These data were taken in similar parts of the solar cycle (June/July, 1998),
and we expect that the solar modulation parameters are
similar between them.
BESS measured the proton flux at an altitude of 37~km
near the geomagnetic north pole
and evaluated the spectrum at the top of the atmosphere,
which was not much affected
by the geomagnetic cutoff.
AMS measured the flux at various positions on Earth
in energies below and above the geomagnetic cutoff
at an altitude close to low-Earth orbit (380~km).
Above the geomagnetic cutoff the two measurements
agree up to about 50 GeV 
and differ less than 10\% between 50 and 100 GeV \citep{Alcaraz2000Proton2}. 
Their data can be modeled well by equation~(2) with 
$\phi=650~{\rm MV}$,
$A=23.9~{\rm c~s^{-1}~m^{-2}~sr^{-1}~MeV^{-1}}$ 
and $a=2.83$.
Other experiments \citep{Seo1991, Boezio1999a, Menn2000}
can also be represented by the model if the solar modulation
parameter $\phi$ is adjusted to an appropriate solar cycle
(see Figure~2).

The flux at a balloon altitude suffers a small attenuation
due to interactions with air.
The nuclear interaction length
in air is $90.0~{\rm g~cm^{-2}}$,
and the probability for vertically-downward
protons to reach the altitude of GLAST BFEM
(atmospheric depth was $3.8~{\rm g~cm^{-2}}$)
is 95.8\%.
For protons of oblique direction, we calculated the
atmospheric depth in line-of-sight by assuming
the trajectory to be straight.
Then the effective atmospheric depth scales as
$\frac{1}{\cos \theta}$ down to $\cos \theta = 0.2$ and is
$138~{\rm g~cm^{-2}}$ for the horizontal direction, 
by assuming the scale height of the air 
above the altitude of the payload (at $\sim 38~{\rm km}$)
to be 7.6~km, a typical value in the U.S. Standard Atmosphere.
Below the horizon the flux is assumed to be 0.
Our primary proton spectral model function
(equation~6 with 4\% air attenuation) with adjustment for the 
geomagnetic cutoff is shown in Figure~3 with AMS data
taken at similar geomagnetic latitude region.

In energy ranges below the geomagnetic cutoff, 
there are secondary particles produced in the Earth's atmosphere by 
interactions of primaries.
The secondary proton spectra at satellite altitude was measured by
AMS in a wide geomagnetic latitude region, 
showing strong dependence on the geomagnetic cutoff.
We represent their data, region by region, by a
broken power-law or cutoff power-law model.
The formula of the broken power-law is,
\begin{equation}
\begin{array}{ll}
F_{0} \times  
\left( \frac{E_{\rm k}}{\rm 100~MeV} \right)^{-a}
& ({\rm 100~MeV} \le E_{\rm k} \le E_{\rm bk}) \\
F_{0} \times 
\left( \frac{E_{\rm bk}}{\rm 100~MeV} \right)^{-a}
\times \left( \frac{E_{\rm k}}{E_{\rm bk}} \right)^{-b}
& (E_{\rm bk} \le E_{\rm k}) \\
\end{array}
\quad,
\end{equation}
and that of the cutoff power-law model is
\begin{equation}
\begin{array}{ll}
F_{1} \times  
\left( \frac{E_{\rm k}}{\rm GeV} \right)^{-a} \times
e^{-\left( \frac{E_{\rm k}}{E_{\rm c}} \right)^{-a+1}}
& ({\rm 100~MeV} \le E_{\rm k}) \\
\end{array}
\quad.
\end{equation}
Below 100~MeV, we found no data and assume
a power-law spectrum with index of -1:
\begin{equation}
\begin{array}{ll}
F_{0} \times  
\left( \frac{E_{\rm k}}{\rm 100~MeV} \right)^{-1}
& ({\rm 10~MeV} \le E_{\rm k} \le {\rm 100~MeV}) \\
\end{array}
\quad.
\end{equation}
Below a few $\times$ 10~MeV, there is no probability for protons
to trigger BFEM/LAT, hence the spectral model 
below 100~MeV (equation~9)
does not affect the data very much.
We tabulated the model parameters to describe the AMS data
in Table~1.
In low geomagnetic latitude regions
($\theta_{\rm M} \le 0.6$) 
the downward and upward spectra
are identical, indicating that protons are confined locally 
by the geomagnetic field.
The particle direction is expected to be randomized,
and we adopt a uniform angular distribution.
For $\theta_{\rm M} \ge 0.6$,
we assume a uniform zenith angle dependence
in the upper and lower hemispheres, for we lack  experimental data
about angular dependence.
Flux in the horizontal direction
is uncertain, and we leave it discontinuous.
Analysis of the BFEM data ($\S~5$) showed that 
horizontal particles ($\theta = 90^{\circ} \pm 20^{\circ}$)
have little chance to trigger the BFEM.
The 3-in-a-row trigger is improbable
for the narrow geometry of this single tower.

The secondary proton spectrum at balloon altitude
is not necessarily the same as that in low-Earth orbit,
for the Larmor radius of 
a typical secondary proton is
much smaller than the satellite altitude:
for example, the radius of a 1~GeV proton is 100~km 
for average Earth magnetic field of 0.3~gauss.
Hence we must refer to existing balloon measurements.
The spectra measured at $R_{\rm cut} \sim 4.5$~GV
at atmospheric depth similar to our experiment
($\rm 3.8\ g\ cm^{-2}$) are collected in Figure~3
\citep{Verma1967, Pennypacker1973, Abe2003}.
Although the flux differs from experiment to experiment,
vertically-downward flux at
${\rm \sim 4\ g\ cm^{-2}}$ is much higher than that measured
by AMS (at 380~km): they could be represented as
\begin{equation}
0.17 \times 
\left( \frac{E_{\rm k}}{\rm 100~MeV} \right)^{-1.0}
\ {\rm c\ s^{-1}\ m^{-2}\ sr^{-1}\ MeV^{-1}}
\quad,
\end{equation}
in 10~MeV--4~GeV.
Above the geomagnetic cutoff, we assume that
the secondary proton flux follows
a power-law function of 
$\left( \frac{E_{\rm k}}{\rm GeV} \right)^{-2.83}$, 
i.e., that of the primary.
Note that the downward flux is known to be proportional to
the atmospheric depth \citep{Verma1967, Abe2003},
and balloon data in Figure~3 were already scaled to $3.8~{\rm g~cm^{-2}}$.
On the other hand, upward flux does not depend on the atmospheric depth
very much (e.g., Verma 1967) and 
we modeled the data with the following formulas.
They are
\begin{equation}
0.17 \times 
\left( \frac{E_{\rm k}}{\rm 100~MeV} \right)^{-1.6}
\ {\rm c\ s^{-1}\ m^{-2}\ sr^{-1}\ MeV^{-1}}
\quad,
\end{equation}
for energy above 100~MeV and 
\begin{equation}
0.17 \times  
\left( \frac{E_{\rm k}}{\rm 100~MeV} \right)^{-1.0}
\ {\rm c\ s^{-1}\ m^{-2}\ sr^{-1}\ MeV^{-1}}
\quad,
\end{equation}
for energy below 100~MeV. 
As can be seen from Figure~3,
secondary fluxes measured by balloon experiments
are mutually inconsistent within a factor of 2.
Hence our model 
inherits this much uncertainty.

The angular distribution of the secondary cosmic-ray flux 
has been measured in several
balloon and rocket experiments.
While substantial build-up of the flux near the Earth horizon
has been reported in several experiments
\citep{Allen1950a, Allen1950b, Singer1950}, 
its functional form is not well determined.
For small zenith angles, the secondary proton downward flux is expected to be
proportional to the atmospheric depth in 
line-of-sight, and we multiply 
the flux
by $\frac{1}{\cos \theta}$. The flux is expected to be saturated 
near the horizon and we use a constant factor of 2 for
$60^{\circ} \le \theta \le 90^{\circ}$.
For upward flux, we assume the same angular dependence 
as a function of nadir angle instead of zenith angle.
Again the flux disconnects at horizon due to uncertainty.

\subsection{Alpha Particle Flux}
We adopted the AMS and BESS data to model the alpha particle spectrum.
The two experiments are
consistent with each other to $\sim$10~\%
\citep{Alcaraz2000Helium}, giving 
$A=1.50\ {\rm c\ s^{-1}\ m^{-2}\ sr^{-1}\ MeV^{-1}}$,
$a=2.77$ and $\phi=650~{\rm MV}$ for equation~2.
Other experiments 
\citep{Seo1991, Boezio1999a, Menn2000}
can also be represented well by adjusting the solar
modulation parameter $\phi$ (Figure~4).
We did not take the secondary component into account, 
for the flux is 
more than three orders of magnitude lower than the primary,
as shown by \citet{Alcaraz2000Helium}.
For the GLAST balloon experiment, we used equation~3
with ${\rm \phi=1100\ MV}$ and $R_{\rm cut}=4.46\ {\rm GV}$.
The zenith angle dependence and the attenuation by air
are assumed to be the same as that of proton primaries.

\subsection{Electron and Positron Flux}
The fluxes of cosmic-ray electrons and positrons are
modeled in 3 components: the primary, the secondary downward
and the secondary upward fluxes.

Unlike proton and alpha primaries,
no single experiment has measured cosmic primary electron
and positron fluxes over a wide energy range.
We referred to a compilation of measurements  
by Webber \citep{Webber1983, HEAP}
and used the power-law fit given there, i.e., 
\begin{equation}
Unmod(E_{\rm k})
=0.7\times \left( \frac{R(E_{\rm k})}{\rm GV}\right)^{-3.3}
\ {\rm c\ s^{-1}\ m^{-2}\ sr^{-1}\ MeV^{-1}}~.
\end{equation}
A recent review of measurements by \citet{Moskalenko1998}
gave a consistent spectrum above 10~GeV.
The positively charged fraction, interpreted as
$\frac{e^{+}}{e^{-} + e^{+}}$,
has been measured by several experiments. Among them, 
\citet{Golden1994}
obtained the fraction to be a constant,
$0.078 \pm 0.016$, between 5~GeV and 50~GeV.
We adopted their results and modeled the
interstellar electron spectrum 
to be equation~1 with 
$A=0.65\ {\rm c\ s^{-1}\ m^{-2}\ sr^{-1}\ MeV^{-1}}$ 
and $a=3.3$. 
The spectrum of the primary positrons takes the same form
except for the normalization factor.
Like proton and alpha primaries,
the angular distribution is assumed to be 
uniform and independent of zenith angle $\theta$
for $\theta \le \theta_{\rm c}$ and
zero for $\theta \ge \theta_{\rm c}$, where $\theta=0^{\circ}$
corresponds to particle going toward the nadir.
The solar and geomagnetic modulation parameters
for the GLAST Balloon Experiment are
$\phi=1100\ {\rm MV}$ and $R_{\rm cut}=4.46\ {\rm GV}$.
To model the $e^{-}$/$e^{+}$ primary flux for our experiment,
we also need to take into account the
energy loss due to the ionization and bremsstrahlung in the atmosphere.
For vertically downward particles it is calculated to be
$e^{-\frac{\rm 3.8~g~cm^{-2}}{\rm 36.6~g~cm^{-2}}} = 0.90$,
where ${\rm 36.6~g~cm^{-2}}$ is the radiation length of air.
This energy loss gives the same spectral shape but smaller normalization
by a factor of 1.42.
We calculated the atmospheric depth in the line-of-sight 
as a function of zenith angle
in the same way as that for proton and alpha primaries,
and took the energy loss into account.

Secondary cosmic ray electrons and positrons are generated in two
parts, a downward-moving component and an upward-moving one.
To model them at a satellite altitude, we referred to 
the AMS data and represented them with a power-law model:
\begin{equation}
\begin{array}{ll}
F_{0} \times 
\left( \frac{E_{\rm k}}{\rm 100~MeV} \right)^{-a}
& ({\rm 100~MeV} \le E_{\rm k}) \\
\end{array}
\quad,
\end{equation}
or a broken power-law model (equation~7), or 
a power-law model with a hump:
\begin{equation}
\begin{array}{ll}
F_{0} \times 
\left( \frac{E_{\rm bk}}{\rm 100~MeV} \right)^{-a} +
F_{1} \times
\left( \frac{E_{\rm k}}{\rm GeV} \right)^{b} \times
e^{-\left( \frac{E_{\rm k}}{E_{\rm c}} \right)^{b+1}}
& ({\rm 100~MeV} \le E_{\rm k}) \\
\end{array}
\quad.
\end{equation}
Below 100~MeV we found no data and use the power-law model
with a spectral index of -1, i.e., equation~9.
Again there is no probability for $e^{-}/e^{+}$
to cause trigger below 10~MeV.
The model functions are tabulated in Table~2:
in near geomagnetic equator, $e^{+}$ flux is much higher
than that of $e^{-}$.
We use the same flux models for downward and upward component,
since AMS did not detect significant difference between them.
As for proton secondaries,
a uniform angular distribution is assumed
in upper and lower hemispheres.

Most of the existing balloon measurements of
secondary electron flux at around $R_{\rm cut}=4.5\ {\rm GV}$ 
\citep{Verma1967, Israel1969, HEAT1}
were rather old and inconsistent with each other,
as shown in Figure~5.
However, they consistently show much higher and flatter
spectra than that of AMS.
We therefore referred to balloon measurements and adopted 
a higher and flatter spectral model 
for the GLAST Balloon Experiment.
The vertically downward spectrum is
\begin{equation}
0.41 \times
\left( \frac{E_{\rm k}}{\rm 100~MeV} \right)^{-2.1}
\ {\rm c\ s^{-1}\ m^{-2}\ sr^{-1}\ MeV^{-1}}
\quad,
\end{equation}
for 100~MeV--4~GeV and
\begin{equation}
0.41 \times
\left( \frac{E_{\rm k}}{\rm 100~MeV} \right)^{-0.5}
\ {\rm c\ s^{-1}\ m^{-2}\ sr^{-1}\ MeV^{-1}}
\quad,
\end{equation}
in energy 10--100~MeV.
The flux above the geomagnetic cutoff is
expected to follow the functional form of 
proton primaries ($E_{\rm k}^{-2.83}$).
Upward flux has been measured by \citet{HEAT2}
in 1.0--2.4~GeV
and is similar to
that of downward,
and we used the same spectral model for both downward and
upward for simplicity.
We did not find data at large zenith angle,
and assume the same angular distribution as that
for proton secondaries.

Although AMS \citep{Alcaraz2000Lepton}
observed an overabundance of positrons 
relative to electrons in the
low geomagnetic latitude region,
fluxes of these two particles are 
almost identical in
the region of $\theta_{\rm M} \ge 0.6$, as shown in Table~2. 
A recent balloon measurement by \citet{HEAT1}
also showed that the positron fraction is nearly 50\% at 
$R_{\rm cut}=4.5\ {\rm GV}$. 
We therefore used the same spectral shape and flux
for positron secondaries.
Note that the data points of balloon experiments referenced,
where only the flux of $e^{-} + e^{+}$ is given,
have been divided by 2 to convert to the $e^{-}$ flux
in Figure~5.

\subsection{Gamma-ray Flux}
As for particles at balloon altitudes, the gamma-ray flux 
consists of primary and
secondary components. The primary component is of
extraterrestrial origin
and the intensity is attenuated by air at balloon altitude.
Secondaries are produced by the interaction between cosmic-ray particles
and the atmospheric molecules, and the intensity of the downward component
depends on the residual atmospheric depth.

\subsubsection{Primary Gamma-ray Flux}
We modeled the extragalactic diffuse gamma-rays 
measured by several gamma-ray satellites, as compiled by \citet{Sreekumar1998}.
Above 1~MeV, the spectrum can be expressed by a power-law function of
\begin{equation}
40.0 \times \left( \frac{E_{\rm k}}{\rm MeV} \right)^{-2.15}
~{\rm c~s^{-1}~m^{-2}~sr^{-1}~MeV^{-1}}~.
\end{equation}
The zenith-angle dependence is the same as those of
proton/alpha/$e^{\pm}$ primaries.
For energies where GLAST BFEM has sensitivity for gamma rays
($E_{k} \ge 1~{\rm MeV}$),
the primary flux is lower than that of 
secondary downward gamma by more than an order of magnitude. 
We therefore do not take into account the 
gamma primary component in the following analysis 
of BFEM data (\S~5).

\subsubsection{Secondary Gamma-ray Flux}

Secondary gamma-rays are produced either
by interactions of cosmic-ray hadrons or 
bremsstrahlung of $e^{\pm}$ with nuclei in the
atmosphere.
The spectrum has been measured by several satellite and balloon
experiments
\citep{Thompson1974, Imhof1976, Ryan1979, Kur'yan1979, Shonfelder1980, GammaReview}.
For residual atmosphere less than $\rm 100\ g\ cm^{-2}$, 
the vertically-downward gamma-ray flux is
known to be proportional to the atmospheric depth, 
whereas the upward flux remains 
almost constant \citep{Thompson1974}.
Observations at Palestine of the vertically downward spectrum
at the atmospheric depth of ${\rm 2.5~g~cm^{-2}}$
are compiled in \citet{Shonfelder1980} and \citet{GammaReview},
showing  consistent results. 
We multiplied their data by $\frac{3.8}{2.5}$
to scale them to the residual atmosphere of 
$\rm 3.8\ g\ cm^{-2}$,
the value for the GLAST Balloon Experiment, and modeled 
the spectrum as follows: 
For 1-1000 MeV, it is
\begin{equation}
250 \times \left( \frac{E_{\rm k}}{\rm MeV} \right)^{-1.7}+
1.14 \times 10^{5} \times 
\left( \frac{E_{\rm k}}{\rm MeV} \right)^{-2.5} 
\times e^{- \left( \frac{E_{\rm k}}{E_{\rm c}}\right)^{-1.5}}
\ {\rm c\ s^{-1}\ m^{-2}\ sr^{-1}\ MeV^{-1}}
\quad,
\end{equation}
where $E_{\rm c}=120\ {\rm MeV}$. For 1--100~GeV,
the spectrum is
\begin{equation}
2.15 \times 10^{4} \times \left( \frac{E_{\rm k}}{\rm MeV} \right)^{-2.2}
\ {\rm c\ s^{-1}\ m^{-2}\ sr^{-1}\ MeV^{-1}}
\quad.
\end{equation}
The composite spectrum is shown with reference data in Figure~6.
Note that we do not expect this component in satellite orbit.

For the vertically-upward component, 
only a few data are available
\citep{Thompson1974, Imhof1976, Ryan1979, Kur'yan1979}
as  collected in Figure~7.
We found data taken at Palestine, Texas 
($R_{\rm cut} \sim 4.5\ {\rm GV}$) down to a few MeV, 
whereas we found none below a few MeV:
so we substituted with data by \citet{Imhof1976}
at $R_{\rm cut}=3\ {\rm GV}$. 
The flux depends on
cut-off rigidity, as discussed in 
\citet{Kur'yan1979}:
they measured the flux of vertically-upward gammas above 80~MeV,
for which GLAST LAT/BFEM has high sensitivity, in $R_{\rm cut}$ of
4--17.5~GeV, and showed that 
the flux anticorrelates the cutoff rigidity as
${R_{\rm cut}}^{-1.13}$.
\citet{Thompson1981} confirmed the dependence 
by comparing the gamma-ray intensity
above 35~MeV at equator and at $R_{\rm cut} \sim 4.5~{\rm GeV}$.
We adopted their results and
scaled the flux at $R_{\rm cut}=3\ {\rm GV}$ by a factor of
$\left( \frac{4.5}{3.0} \right)^{-1.13}=0.63$
to take the dependence into account. 
The composite spectrum for vertically upward flux
seems to be represented well by two power-law functions:
for 1--20~MeV, it is
\begin{equation}
1010 \times 
\left( \frac{E_{\rm k}}{\rm MeV} \right)^{-1.34}
\ {\rm c\ s^{-1}\ m^{-2}\ sr^{-1}\ MeV^{-1}}
\quad,
\end{equation}
and for 20--1000~MeV,
\begin{equation}
7290 \times 
\left( \frac{E_{\rm k}}{\rm MeV} \right)^{-2.0}
\ {\rm c\ s^{-1}\ m^{-2}\ sr^{-1}\ MeV^{-1}}
\quad.
\end{equation}
Above 1 GeV we could not find data and 
adopted the same spectral index as that of downward gamma rays
($E^{-2.2}$). 
Then, the spectrum in 1--1000~GeV is expressed as
\begin{equation}
2.9 \times 10^{4} \times  
\left( \frac{E_{\rm k}}{\rm MeV} \right)^{-2.2}
\ {\rm c\ s^{-1}\ m^{-2}\ sr^{-1}\ MeV^{-1}}
\quad.
\end{equation}
The spectral model with reference data
points is given in Figure~7. As can be seen from the figure,
observations are very scarce and give inconsistent
results with each other. 
We regard that the uncertainty is a factor of 3.

\subsubsection{Zenith Angle Dependence of Secondary Gamma Rays}
Angular dependence of the secondary gamma-ray flux has been measured by
some authors \citep{Thompson1974, Shonfelder1977}
and is known to depend strongly on zenith
angle. We constructed the model function referring to 
\citet{Shonfelder1977}, where they
measured the flux in 1.5--10 MeV region at 
$\rm 2.5\ {\rm g\ cm^{-2}}$. 
We multiplied their data between 
$0^{\circ} \le \theta < 90^{\circ}$ (i.e., downward component) by
3.8/2.5 to correct the atmospheric depth,
and expressed the relative flux as
\begin{equation}
\begin{array}{ll}
\frac{1}{\cos \theta} & (0^{\circ} \le \theta \le 60^{\circ}) \\
0.367 \times e^{1.618 \theta} & (60^{\circ} \le \theta \le 90^{\circ}) \\
8.71 \times 10^{-3} \times e^{4.0 \theta} & (90^{\circ} \le \theta \le 115^{\circ}) \\
25760 \times e^{-3.424 \theta} & (115^{\circ} \le \theta \le 140^{\circ}) \\
6 & (140^{\circ} \le \theta \le 180^{\circ}) \\
\end{array}
\quad .
\end{equation}
Here, $\theta$ in model functions is given in radians.
True angular dependence of the flux and spectral shape could be 
correlated with each other. For simplicity, we
assumed that the functions above represents the zenith-angle dependence at
3~MeV and the spectral shape is independent of zenith angle
except for the difference between downward and upward.
Our model function and referenced data are given in Figure~8 with a typical measurement
error, indicating that the uncertainty is large, up to a factor of 2.

\subsection{Fluxes of $\mu^{\pm}$}
Due to its short lifetime,
we can neglect any primary muon
component. The spectrum of atmospheric muons is reported by 
\citet{Boezio2000},
where they measured the flux of downward $\mu^{\pm}$  
in 0.1-10 GeV band. 
At float altitude (residual atmospheric depth is $3.9~{\rm g~cm^{-2}}$),
the $\mu^{-}$ spectrum can be expressed as
\begin{equation}
6.5 \times 10^{-3} \times
\left( \frac{E_{\rm k}}{\rm GeV} \right)^{-2.2}
\times e^{-\left( \frac{E_{\rm k}}{\rm 0.43\ GeV} \right)^{-1.5}}
\ {\rm c\ s^{-1}\ m^{-2}\ sr^{-1}\ MeV^{-1}}
\quad,
\end{equation}
for 300~MeV--20~GeV, and as
\begin{equation}
1.65 \times 10^{-2}
\ {\rm c\ s^{-1}\ m^{-2}\ sr^{-1}\ MeV^{-1}}
\quad,
\end{equation}
for 10--300~MeV.
We could not find data for the upward spectrum
and assumed the same function for simplicity.
Data of \citet{Boezio2000}
were obtained at geomagnetic polar region,
whereas GLAST Balloon Experiment was conducted at 
$R_{\rm cut} \sim 4.5$~GV. 
They also 
measured muon spectra at $R_{\rm cut} \sim 4.5$~GV
in energy above 2 GeV \citep{Boezio2003}
and obtained a consistent spectrum.
A recent measurement by \citet{Abe2003} at 
$R_{\rm cut} \sim 4.5~{\rm GeV}$ also gave a consistent spectrum
for energies above 0.5~GeV.
We therefore used the
model functions above (equation~25 and 26)
for the GLAST Balloon Experiment.
The ratio of $\mu^{+}$ to $\mu^{-}$, 
defined as $\frac{\mu^{+}}{\mu^{-}}$, 
is found to be 1.6 in the data of \citet{Boezio1999b}:
so we modeled the $\mu^{+}$ spectrum with the same spectral shape 
but 1.6 times higher normalization. 
As shown by \citet{Boezio1999b} and
\citet{Abe2003}, the downward $\mu^{\pm}$ flux is proportional to the
atmospheric depth. We therefore assumed the same angular dependence
as that of proton secondaries.
Our model functions and reference data
are given in Figure~9.
Note that we do not expect muons in satellite orbit:
even for 10~GeV muons the life time is $180~\mu s$ in the laboratory frame
and they can travel only 50~km.

\section{GLAST Balloon Flight}
After the integration and test at Stanford Linear Accelerator
Center and Goddard Space Flight Center, the instrument was
shipped to the National Scientific Balloon Facility (NSBF)
in Palestine, Texas.
On August 4, 2001, the BFEM was
successfully launched.
The balloon carried the instrument to an altitude of 38~km
in about two hours, and achieved about three hours of level flight
until the balloon reached the limit of telemetry from the NSBF.
The atmospheric depth at the float altitude was about 
${\rm 3.8\ g\ cm^{-2}}$.
Due to a small leak of the pressure vessel, the internal pressure went down to
0.14 atmosphere and we failed to record all the data to hard disk.
However, a sample of triggered data was continuously obtained
via telemetry with about 12~Hz out of 
a total trigger rate of 500~Hz.
More than $10^{5}$ events were recorded during the level flight and
showed that all detectors and DAQ system kept working
in a space-like high-counting environment.
The data obtained allow us to study 
the cosmic-ray background environment.
More details of the preparation and the flight of the BFEM
were described in \citet{Thompson2002}.
In the next section, we compare the simulation prediction
with the observed data.

\section{Comparison of Data and Simulation}
After the flight we found two issues 
to be taken into account in data analysis.
First, three layers out of 26,
i.e., 7th layer, 15th layer and 16th layer, were found
not to participate in the trigger, 
although hits in these layers were
recorded normally when the trigger condition was satisfied.
We took this effect into account in the Geant4 simulation.
Fortunately, three or more consecutive pairs of x and y layer
(which is the minimum to cause a trigger) worked
in the lower part of the TKR (1st to 6th layer),
in the middle part of the TKR (9th to 14th layer),
and in the upper part of the TKR (17th to 26th layer),
and the loss of trigger efficiency due to this failure
was only $\sim$ 30\%.
The second issue was related to the event sampling in telemetry. 
The sampling unintentionally had a bias toward events 
with smaller size (primarily number of hits in the TKR) in some data packets.
To select unbiased events we required that the
packet contain exactly 10 events; 
a detailed study showed such packets 
to have an unbiased sample within 10\%. 
This selection reduces the number of events,
but we still have 36k events to be used for data analysis.

In order to validate the cosmic-ray flux model and 
the Monte-Carlo simulator described in \S~2 and \S~3, we compared 
the simulation prediction with the data.
The observed trigger rate was about 500~Hz
during the level flight and is in good agreement with
the simulation prediction.
For further study, we divided events into two classes,
``charged events'' and ``neutral events.''
A charged event is one in which any one of 
13 ACD tiles shows energy deposition above 0.38~MeV 
(corresponds to 0.2 of the mean energy deposited in an ACD scintillator 
by a minimum ionizing particle, a MIP).
The rates of charged and neutral events observed were
439~Hz and 58~Hz, respectively, and they are reproduced
by the simulation within 5\%.
The contribution of each particle type to the trigger is summarized in
Table~3, implying that about 70\% of neutral events 
are from atmospheric gamma-rays.
We also compared the count rate of each TKR layer
and summarized the results
in Figure~10.
There, 26 TKR layers are numbered from 0 to 25 in which 
the layer with number 0 is the closest to CAL.
We see jumps in rate between 
layers 15 and 16, and layers 17 and 18
in charged events.
These jumps correspond to the difference of area
covered by SSDs described in \S~2
(area of 3/5 are covered by Si in layers 18--25,
4/5 are covered in layers 16 and 17, and fully covered in 
layers 0--15).
In neutral events we can see three big rate changes above layers 5, 7 and 9,
corresponding to the thick lead converter.
The simulation reproduces not only the trigger rate
but also
these structures very well, validating our
instrument simulator and background flux model.
The differences between the observations 
and model are all less than 10\%, with the largest
difference seen in lower layers (layers 0--5),
where the model overpredicts the rates.
In this region with no converter layers,
the uncertainties of the low-energy
electron, positron and gamma-ray models may affect the results.
The difference could also be because of the uncertainty of passive materials
under the ACD implemented in the simulation.

We also applied the event reconstruction program
that recognizes the pattern to find the tracks in TKR
\citep{Burnett2002} on both
real data and simulation data.
The distribution of the number of reconstructed tracks 
is given in Figure~11, where
the number is two for single track events 
(one in x-layer and the other in y-layer)
and is four for double track events.
The distribution is well reproduced by simulation
for both charged and neutral events.
We also selected the single track events out of the charged events and
compared the goodness-of-fit ($\chi^{2}$)
which evaluates the straightness of the track between data and simulation, 
as shown in Figure~12.
The value of $\chi^{2}$ depends on a particle energy 
and we set it constant (30~MeV) to obtain
a consistent value regardless of whether
the particle hits the CAL or not.
The two peaks in the figure correspond to tracks
with a curved trajectory ($\chi^{2} \ge 0.1$) 
and those with a straight one ($\chi^{2} \le 0.1$).
The former consists of electrons, positrons, 
and gamma rays, whereas the latter is composed of 
protons, alphas and high energy muons. 
The small difference of the lower peak position
between data and simulation could be due to the misalignment
of SSDs. Except for this, the data are well reproduced by simulation,
providing another evidence that the model flux of
each particle type is correct.

Finally we examined the zenith angle distribution of cosmic-ray
particles.
We first selected single track events with a straight track
by requiring $\chi^{2}$ to be less than 0.1, and 
compared the zenith angle distribution between data and
simulation prediction, as shown in Figure~13.
In the reconstruction, particles are assumed to go downward,
therefore $\cos \theta$ ranges from 1.0 (vertically downward)
to 0.0 (horizontal).
Our flux model reproduces the data,
not only for the vertical direction but also for large zenith angle
down to $\cos \theta = 0.3$, within 10\%.
We also applied event selection criteria and studied
the gamma-ray zenith-angle distribution.
The cuts we used here are not optimized for background reduction,
for the number of events is limited.
Instead, we implemented a set of loose cuts to filter out
charged particles and save a
sufficient number of gammas.
To select downward gammas,
we required events with no hit in the ACD,
double tracks in the TKR, and
gaps between two tracks increasing as they go downward 
(downward-opening v-shape trajectory).
For upward gammas, we did not require that the event be neutral,
since the generated electron/positron pair is expected to hit the ACD.
Instead, we required that the two bottommost TKR layers
did not detect a hit and the TKR showed double tracks with
upward-opening v-shape trajectory.
We also set CAL energy threshold at 6~MeV, half of the
energy deposition by MIP crossing one CsI log, to eliminate
events with interaction in CAL.
The results are shown in Figure~14 in which
the bisector of two tracks are assumed to be the direction of gamma.
Data are well reproduced by simulation down to $\cos\theta=0.3$ 
for downward and up to $\cos\theta=-0.3$ for upward,
validating the spectral and angular distribution models for gammas.

\section{Sensitivity of the Model to Assumptions}
The background flux model used to compare with the BFEM data 
involves many parameters, some of which are poorly defined.  
In the course of this investigation, 
we tried a number of alternate assumptions in cases 
where no well-defined observations were available.  
The extent to which such alternatives affected the agreement 
with the data provides an indication of 
how sensitive the model is to such parameters. 
Below we list some of these possibilities 
and describe their influence on the comparison.

$\bullet$ Fluxes of secondary proton, electron, 
and positron were at one time derived directly 
from the AMS measurements. 
Models based on balloon measurements increased the 
trigger rate by about 15\% and improved the agreement of charged events.

$\bullet$ The model function of the secondary upward-moving gamma-ray flux 
were at one time simpler: in 1~MeV to 1~GeV,
it was expressed by a single
power-law function of index of -1.7,
with 1.6 times higher normalization at 1~MeV.
The new model increased the neutral event rate by about 15\%
and improved the match between BFEM data and simulation prediction.

$\bullet$ We once expressed the angular dependence of the
secondary charged particles by $1+0.6 \sin \theta$, an empirical
model given in old rocket measurements 
\citep{Allen1950a, Allen1950b, Singer1950}. The new model did not alter
the trigger rate nor the angular distribution shown by Figure~13
very much.
We adopted our model to take into account the fact that
proton and muon secondary fluxes are proportional to the atmospheric
depth, although the BFEM data alone could not determine which model
is better.

$\bullet$ As described in \S~3, fluxes of secondary
$e^{-}/e^{+}$ and upward gamma are poorly known,
although these particles contributes to the trigger significantly.
Detailed study showed that the main component of
secondary $e^{-}/e^{+}$ which causes the trigger is of energy
of about 100~MeV and that of upward gamma is of energy
of 10--100~MeV, where the particle fluxes are
uncertain by a factor of 2--3.
However, if we change the $e^{-}/e^{+}$ flux by
50\%, the trigger rate of charged event is changed by 
$\sim 50~{\rm c~s^{-1}}$ and disagrees with data.
Similar disagreement will arise in neutral event rate
by adjusting the flux of upward gamma.

With a substantial number of parameters in this model, 
we cannot be certain that all of them are simultaneously optimized. 
What we have found is that specific variations in model parameters 
tend to affect only one aspect of the comparison 
between the model and the BFEM observations.  
The fact that we have obtained reasonable agreement 
with this range of observables - total rate, rate per layer, 
charged component, neutral component, straight tracks, 
curved tracks, and angular distribution - suggests that 
the model is a good approximation to reality.  

\section{Summary}
We have constructed the GLAST BFEM instrument simulator with
a Geant4 toolkit (\S~2)
and the background flux models of abundant components
(proton, alpha, electron, positron, gamma ray and $\mu^{-}/\mu^{+}$)
based on previous measurements and theoretical predictions (\S~3).
The simulation successfully reproduces the observed data within 10\%,
indicating that the flux model is a 
good representation of the cosmic-ray background
at balloon altitude at Palestine, Texas.
Models for the entire low-Earth orbit are also given in \S~3,
where the effect of cutoff rigidity and solar modulation potential 
are parameterized.
They are being used for the simulation of GLAST LAT in orbit
and the study of the background rejection.

The authors would like to thank assistance given by
M. Sj\"ogren, K. Hirano and H. Mizushima in various phases of this work.
They are also grateful to A. Tylka, R. Battiston, Y. Komori, 
A. Yamamoto and T. Yoshida for helpful discussions about
the cosmic-ray background flux. 
They would also like to thank E. do Couto e Silva, H. Tajima,
T. Kotani, R. C. Hartman,
L. Rochester, T. Burnett, R. Dubois, J. Wallace
and GLAST BFEM team and GLAST LAT
collaboration for their support and discussions.
They also would like to thank the staff of National Scientific
Balloon Facility for their excellent support of
launch, flight and recovery.

\clearpage

\begin{table}
\begin{center}
\caption{Model parameters for proton secondary flux at satellite
altitude, based on AMS data.}
\begin{tabular}{cccc}
\tableline\tableline
region & direction & model & 
parameters\tablenotemark{a} \\
\tableline
$0.0 \le \theta_{\rm M} \le 0.2$ & down/upward & cutoff PL 
 & 0.136/0.123/0.155/0.51 \\
$0.2 \le \theta_{\rm M} \le 0.3$ & down/upward & broken PL 
 & 0.1/0.87/600/2.53 \\
$0.3 \le \theta_{\rm M} \le 0.4$ & down/upward & broken PL 
 & 0.1/1.09/600/2.40 \\
$0.4 \le \theta_{\rm M} \le 0.5$ & down/upward & broken PL 
 & 0.1/1.19/600/2.54 \\
$0.5 \le \theta_{\rm M} \le 0.6$ & down/upward & broken PL 
 & 0.1/1.18/400/2.31 \\
$0.6 \le \theta_{\rm M} \le 0.7$ & downward & broken PL 
 & 0.13/1.1/300/2.25 \\
 & upward &  
 & 0.13/1.1/300/2.95 \\
$0.7 \le \theta_{\rm M} \le 0.8$ & downward & broken PL 
 & 0.2/1.5/400/1.85 \\
 & upward &  
 & 0.2/1.5/400/4.16 \\
$0.8 \le \theta_{\rm M} \le 0.9$ & downward & cutoff PL 
 & 0.23/0.017/1.83/0.177 \\
 & upward & broken PL 
 & 0.23/1.53/400/4.68 \\
$0.9 \le \theta_{\rm M} \le 1.0$ & downward & cutoff PL 
 & 0.44/0.037/1.98/0.21 \\
 & upward & broken PL 
 & 0.44/2.25/400/3.09 \\
\tableline
\end{tabular}
\tablenotetext{a}{
$F_{0}$/$a$/$E_{\rm bk}({\rm GeV})$/$b$ 
for the broken power-law (PL) model (equation~7)
and $F_{0}$/$F_{1}$/$a$/$E_{\rm c}$({\rm GeV}) for cutoff PL model
(equation~8).
$F_{0}$ and $F_{1}$ are given in
${\rm c~s^{-1}~m^{-2}~sr^{-1}~MeV^{-1}}$.}
\end{center}
\end{table}

\begin{table}
\begin{center}
\caption{The same as Table~1, but for $e^{-}/e^{+}$
instead of proton.}
\begin{tabular}{cccc}
\tableline\tableline
region & $\frac{e^{+}}{e^{-}}$ ratio & model & 
parameters for $e^{-}$ spectrum \tablenotemark{a} \\
\tableline
$0.0 \le \theta_{\rm M} \le 0.3$ & 3.33 & broken PL &
0.3/2.2/3/4.0 \\
$0.3 \le \theta_{\rm M} \le 0.6$ & 1.66 & PL &
0.3/2.7 \\
$0.6 \le \theta_{\rm M} \le 0.8$ & 1.0 & PL+hump &
0.3/3.3/$2.0 \times 10^{-4}$/1.5/2.3 \\
$0.8 \le \theta_{\rm M} \le 0.9$ & 1.0 & PL+hump &
0.3/3.5/$1.6 \times 10^{-3}$/2.0/1.6 \\
$0.9 \le \theta_{\rm M} \le 1.0$ & 1.0 & PL &
0.3/2.5 \\
\tableline
\end{tabular}
\tablenotetext{a}{$F_{0}$/$a$ for the PL model,
$F_{0}$/$a$/$E_{\rm bk}({\rm GeV})$/$b$ for the broken PL model,
and $F_{0}$/$a$/$F_{1}$/$b$/$E_{\rm c}$({\rm GeV}) for 
PL plus hump. $F_{0}$ and $F_{1}$ are given in
${\rm c~s^{-1}~m^{-2}~sr^{-1}~MeV^{-1}}$.}
\end{center}
\end{table}

\begin{table}[htbp]
\caption{Contribution of each particle type to the trigger rate,
in unit of Hz.}
\small
\begin{flushleft}
\begin{tabular}{lllllllll}
\tableline
particle type & proton & alpha & $e^{-}$/$e^{+}$ & $\gamma$ 2ndary & 
$\gamma$ 2ndary & $\mu^{\pm}$ & total & data \\
& & & & downward & upward & & &\\ 
\tableline
charged events & 216 & 19 & 110 & 12 & 42 & 39 & 438 & 439 \\
neutral events & 4.8 & 0.1 & 12.3 & 16.8 & 26.8 & 0.5 & 61.3 & 58.2 \\ 
\tableline
\end{tabular}
\end{flushleft}
\end{table}

\clearpage

\begin{figure}
\plotone{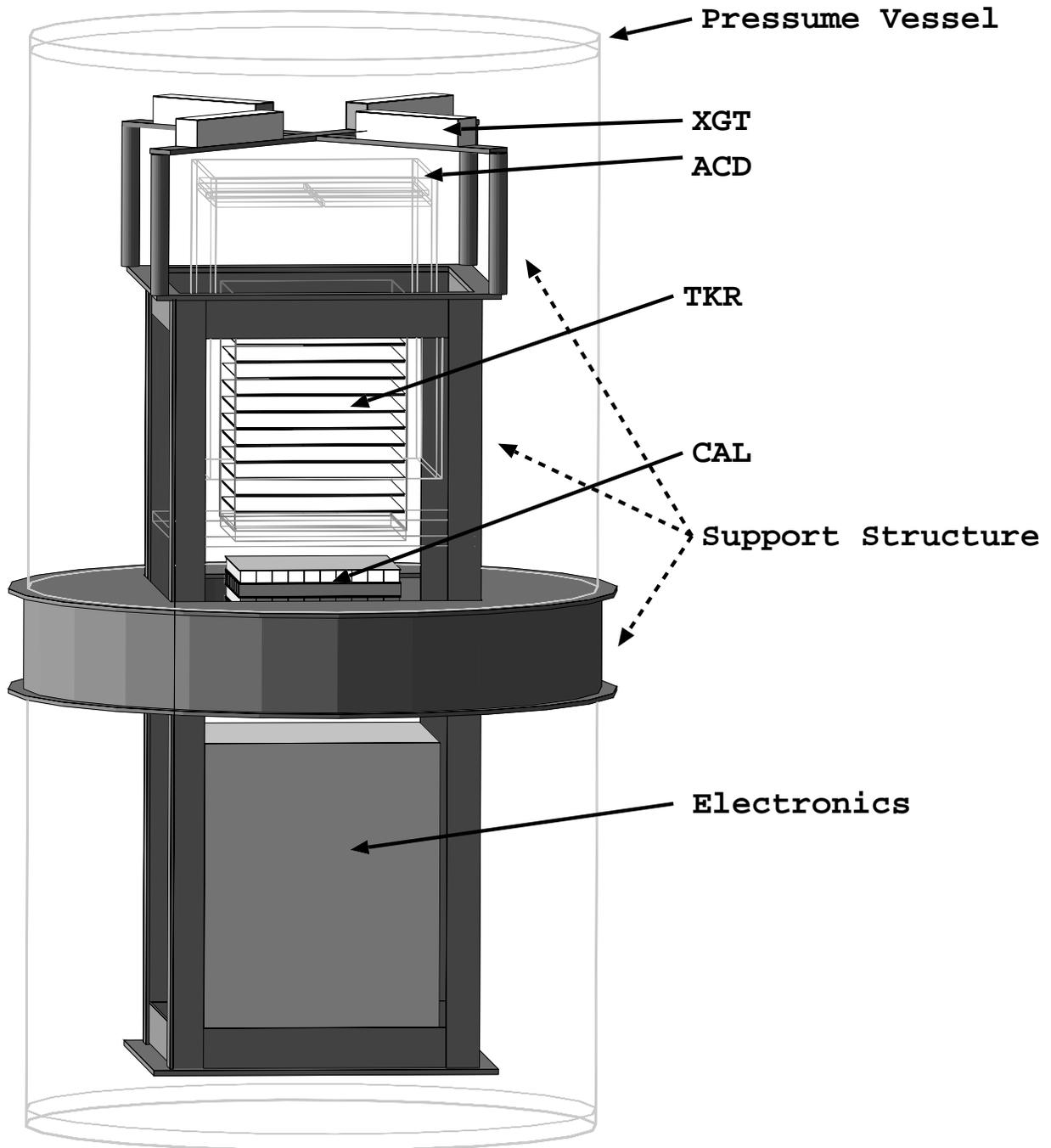}
\caption{
The geometry of the BFEM in Geant4 simulator.
}
\end{figure}

\begin{figure}
\plotone{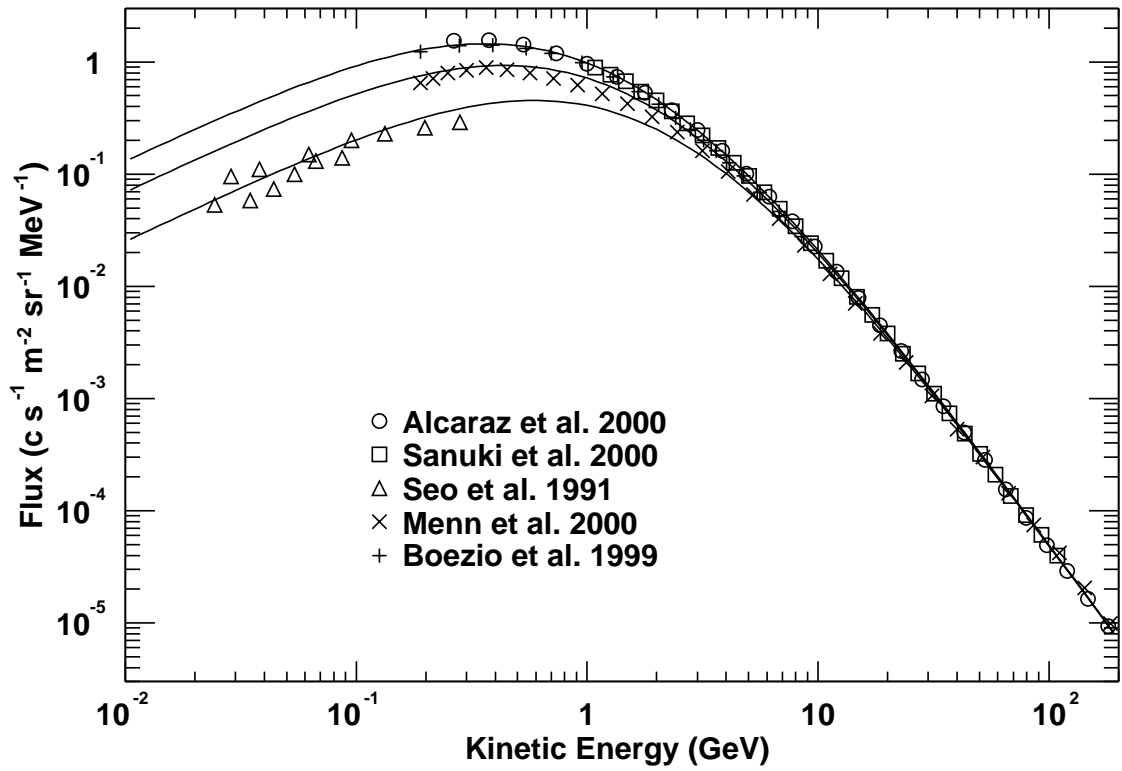}
\caption{
A compilation of measured primary proton spectra in high
geomagnetic latitude region. Three lines are our model functions
(equation~2)
with $\phi$=650~MV (upper line), 800~MV (middle line),
and 1100~MV(lower line).}
\end{figure}

\begin{figure}
\plotone{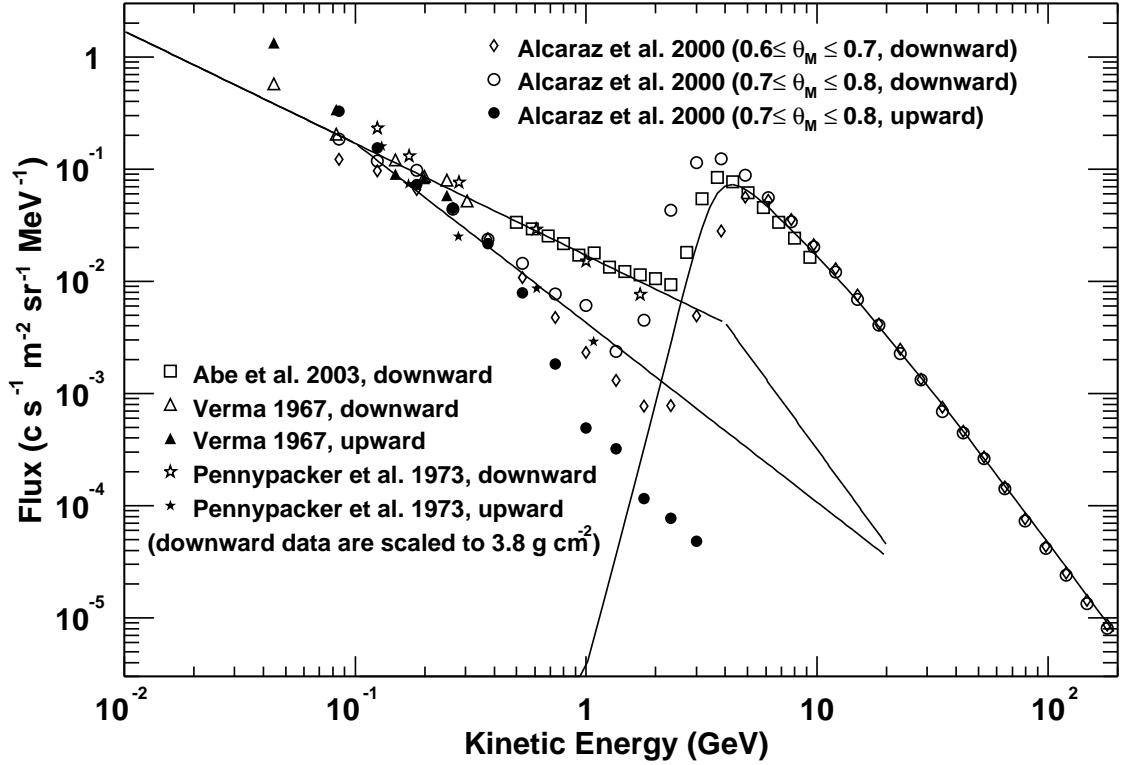}
\caption{The proton vertically downward/upward spectra at
$R_{\rm cut} \sim 4.5$~GV observed by balloon experiments and
AMS. Atmospheric depths are $\rm 4.6\ g\ cm^{-2}$,
$\rm 4\ g\ cm^{-2}$, and $\rm 5\ g\ cm^{-2}$ for data by
Abe et al. (2003), data by Verma (1967), and data by Pennypacker (1973),
respectively. Solid lines are model functions of primary and
secondary proton spectra for GLAST Balloon Experiments.}
\end{figure}

\begin{figure}
\plotone{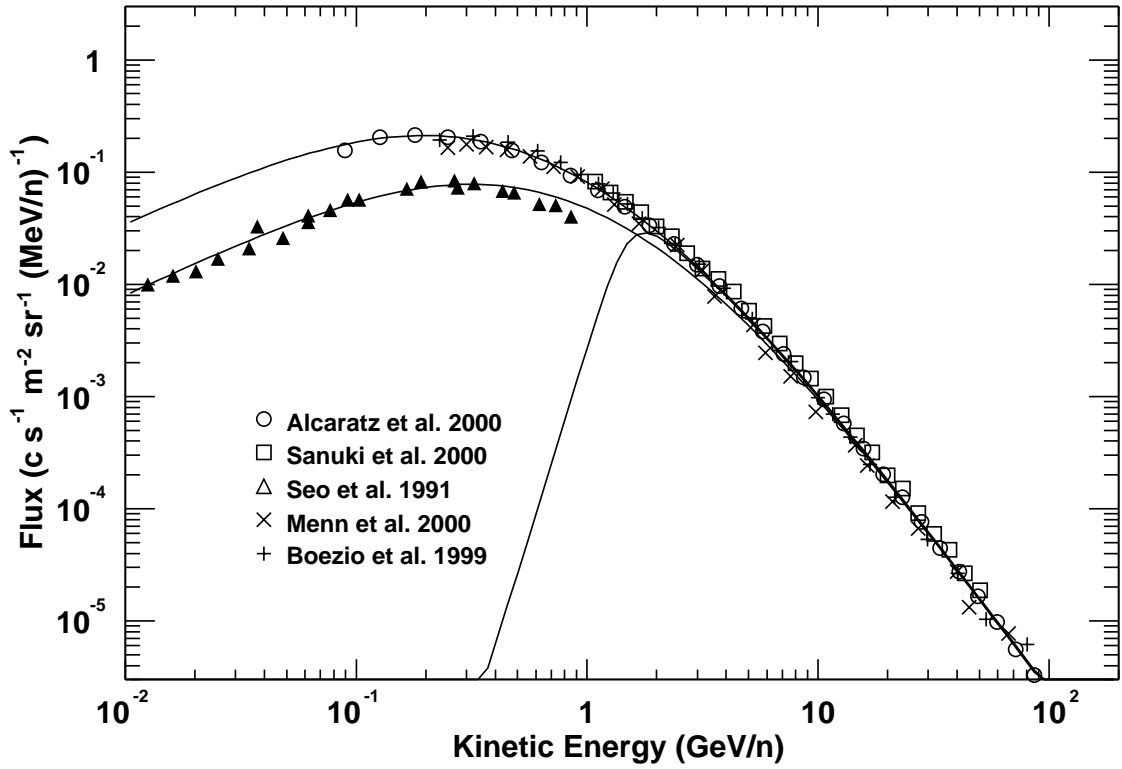}
\caption{The alpha primary flux observed in high geomagnetic region.
Our model functions (equation~2)
of $\phi$=650~MV (upper line) and
1100~MV (lower line) are also shown.
Solid line with a cutoff below a few GeV is the model function
for GLAST BFEM (equation~6).}
\end{figure}

\begin{figure}
\plotone{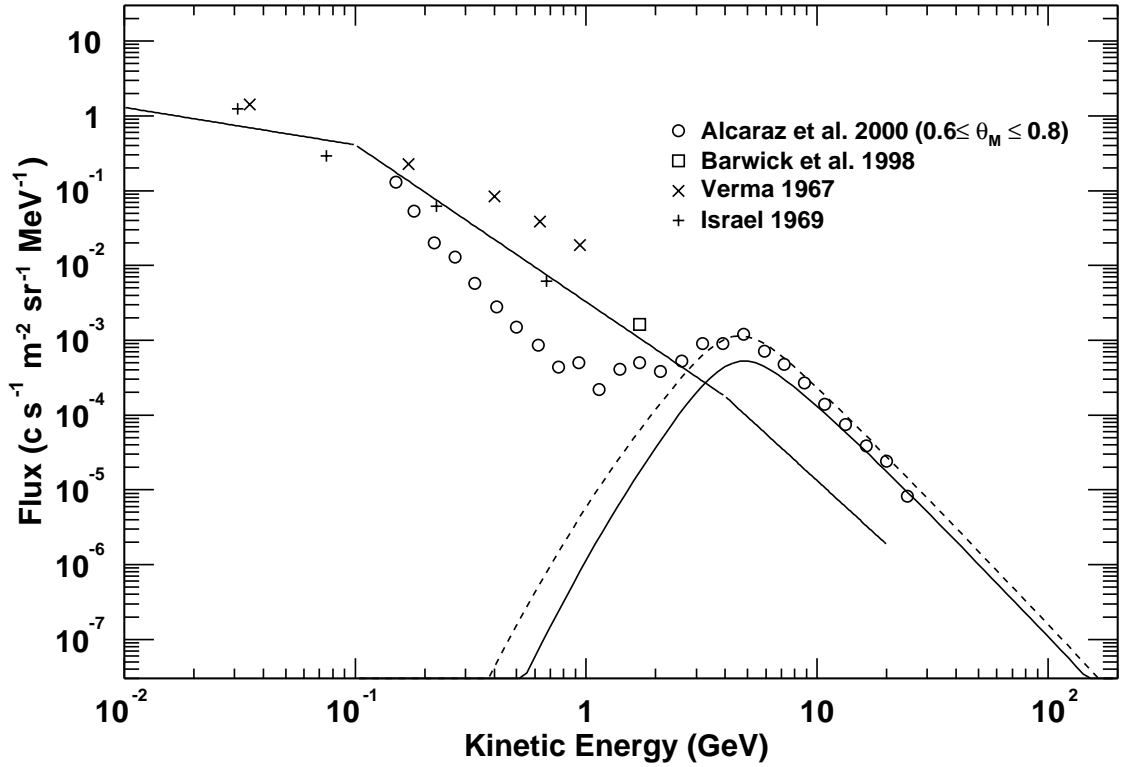}
\caption{Electron vertically downward spectrum at 
$R_{\rm cut} \sim 4.5$~GV measured by balloon experiments and AMS.
Model functions for the GLAST Balloon Experiments are shown by 
solid lines. Dotted line is the primary spectral model with no
air attenuation and of $\phi=650~{\rm MV}$, a value for
AMS observation.}
\end{figure}

\begin{figure}
\plotone{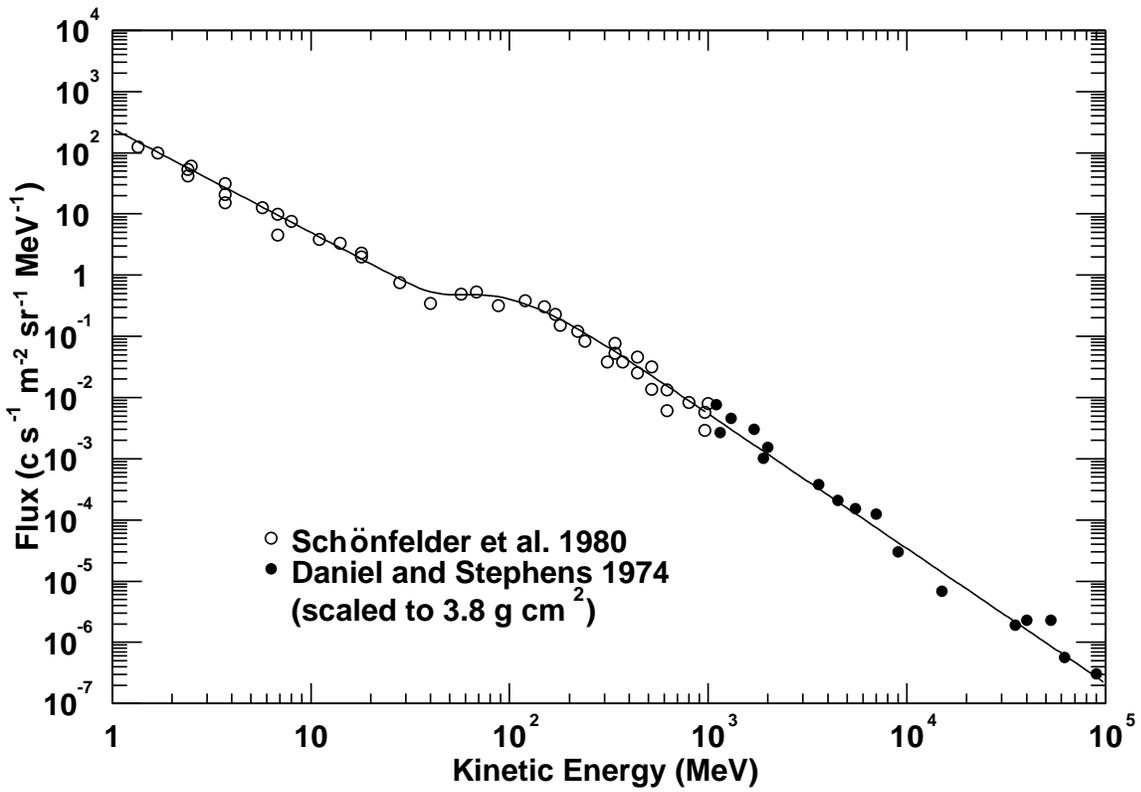}
\caption{Vertically downward atmospheric gamma-ray spectrum.
Spectral model is shown by solid line and reference data by
filled and open circles.}
\end{figure}

\begin{figure}
\plotone{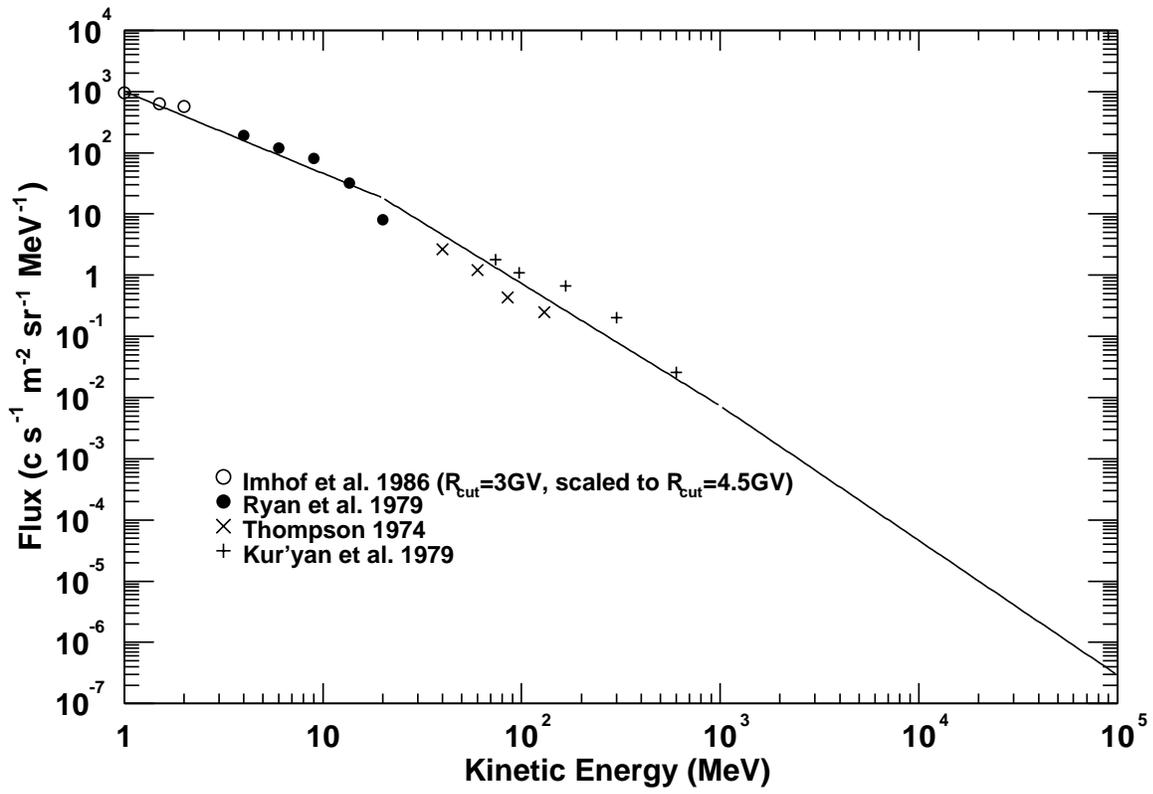}
\caption{Vertically upward gamma-ray spectral model with
reference data.}
\end{figure}

\begin{figure}
\plotone{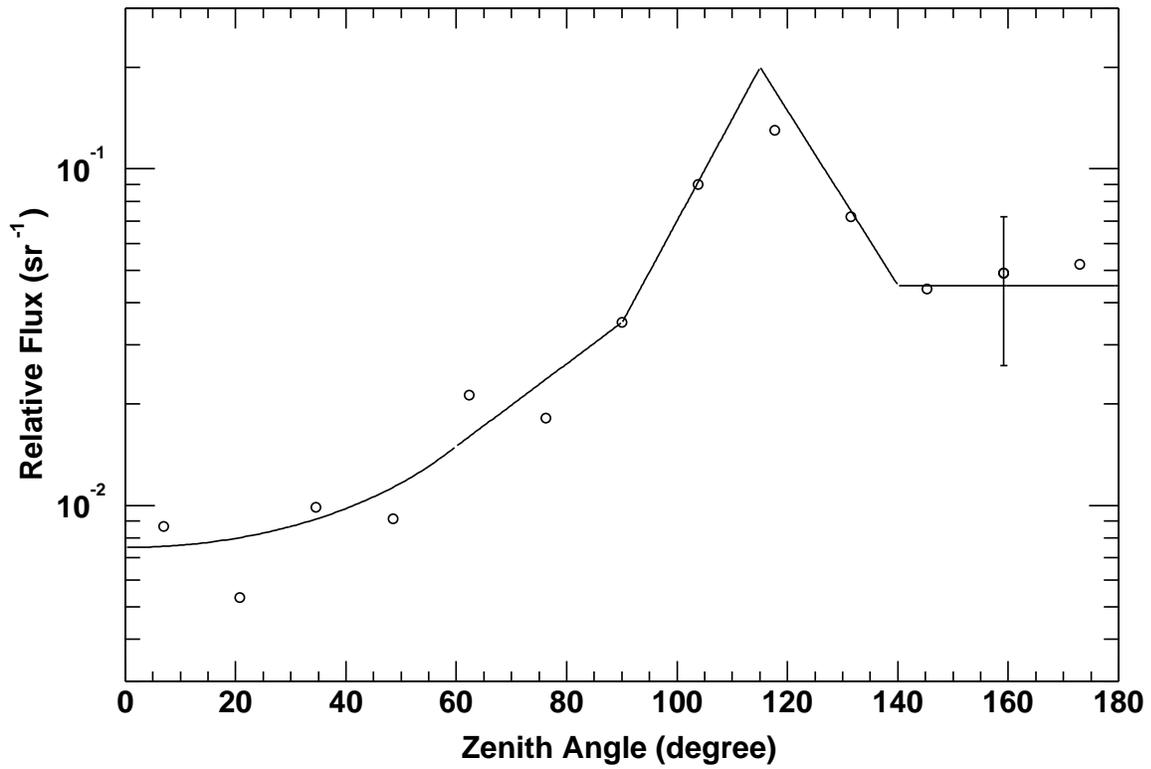}
\caption{Zenith angle dependence of the gamma secondary flux
measured by 
\citet{Shonfelder1977}
in 1.5--10 MeV and
model functions. Data points of downward
are multiplied by a factor of 3.8/2.5 to
correct the atmospheric depth dependence. 
Typical error of the measurement is also shown.}
\end{figure}

\begin{figure}
\plotone{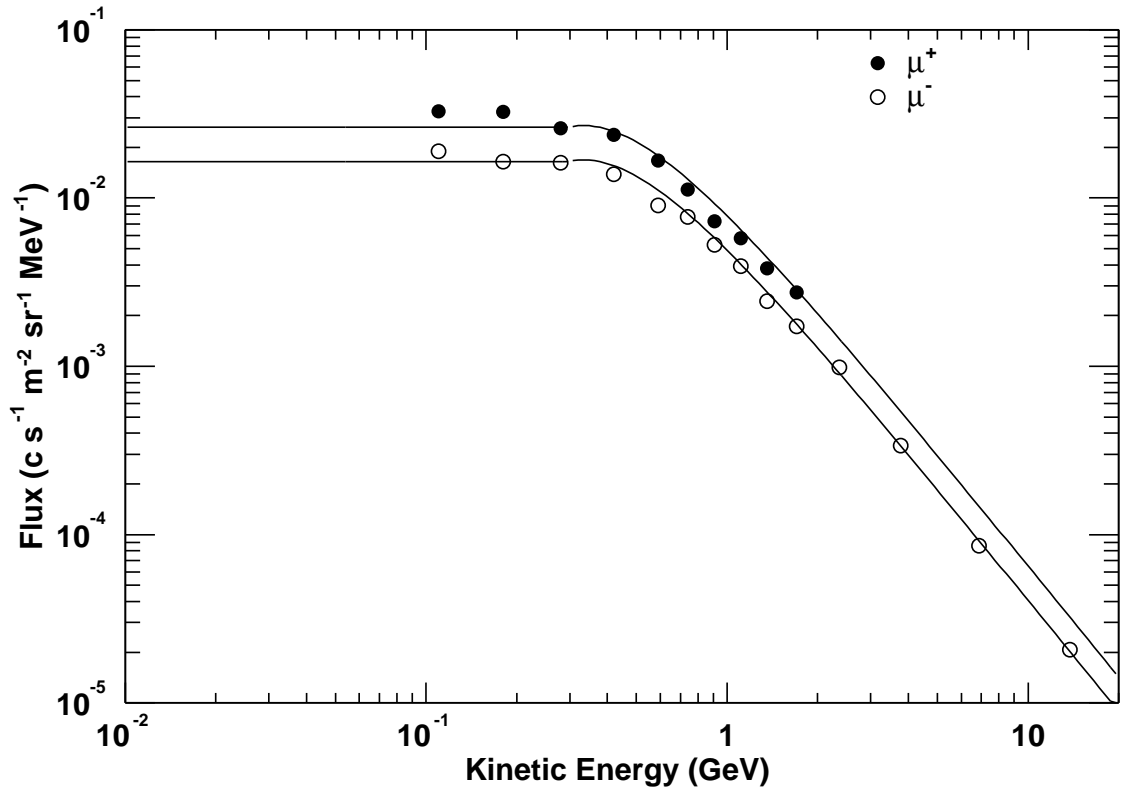}
\caption{Muon model spectra and reference data of 
\citet{Boezio2000}.}
\end{figure}

\begin{figure}
\epsscale{0.8}
\plotone{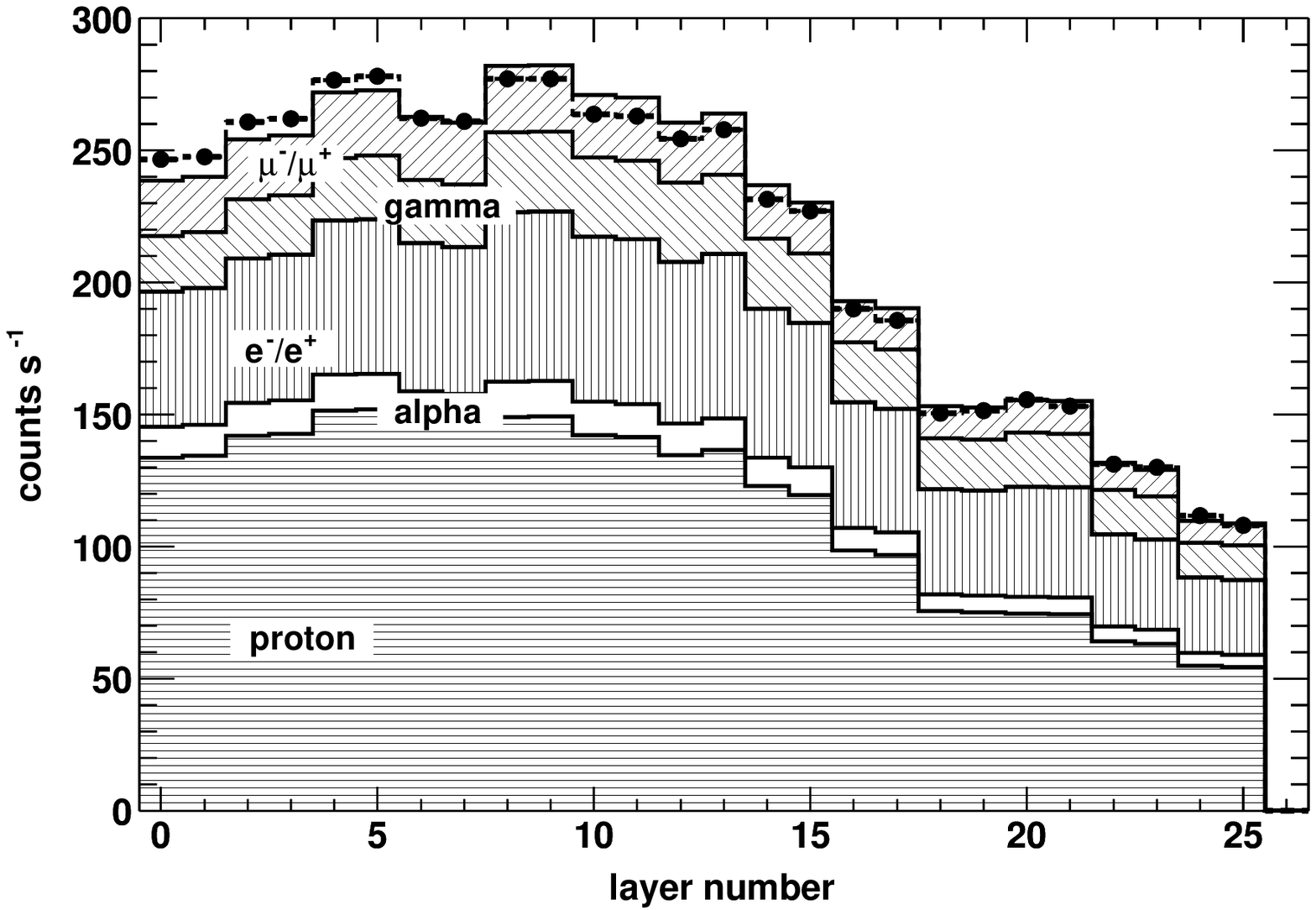}
\plotone{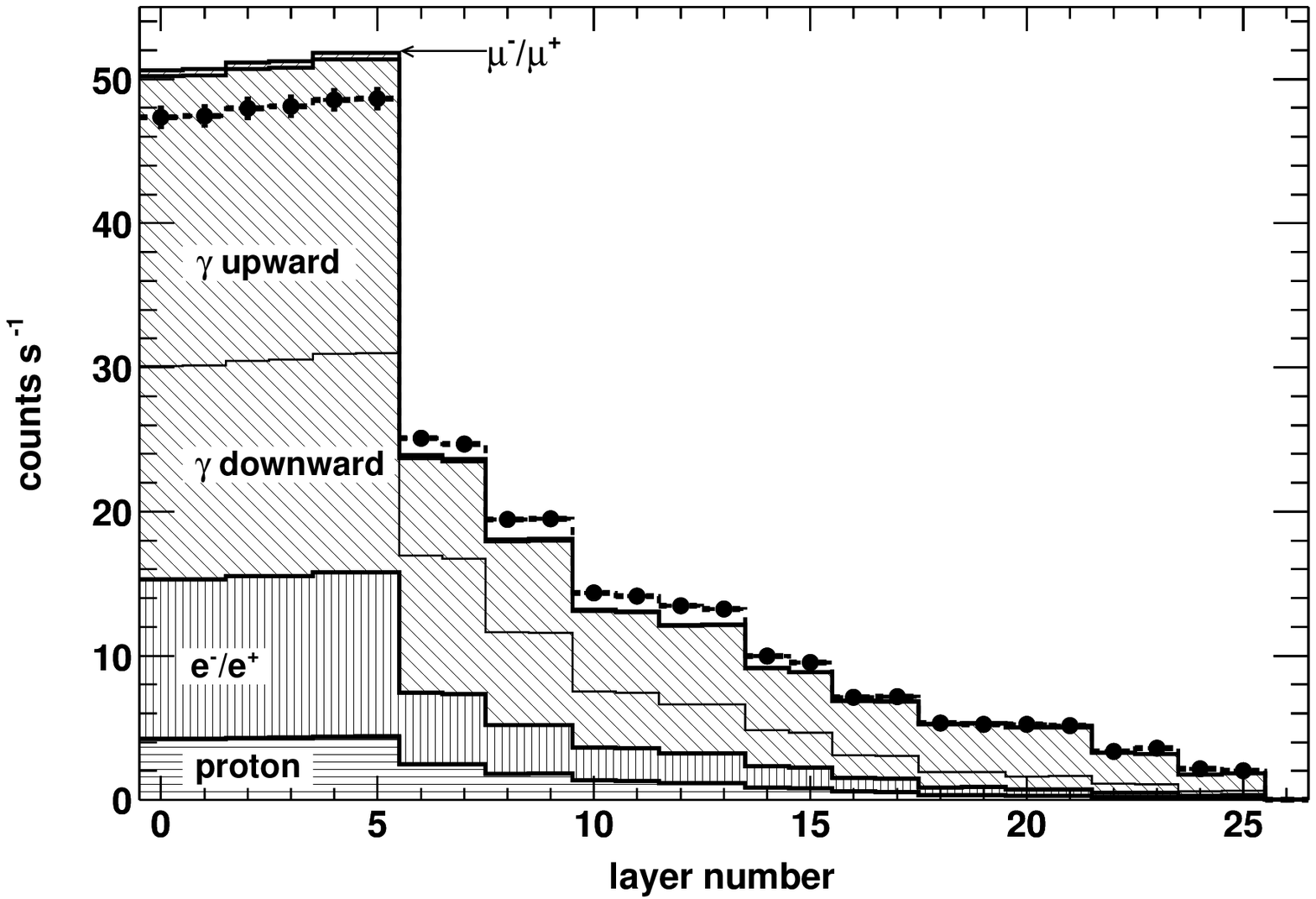}
\caption{Count rate of each layer for charged (top panel)
and neutral (bottom panel) events.
Data-points are shown by circles and simulation prediction by
histogram. Contribution of each particle type is also shown.
For neutral events, contribution of upward gamma
and downward gamma is also presented.}
\end{figure}

\begin{figure}
\epsscale{0.8}
\plotone{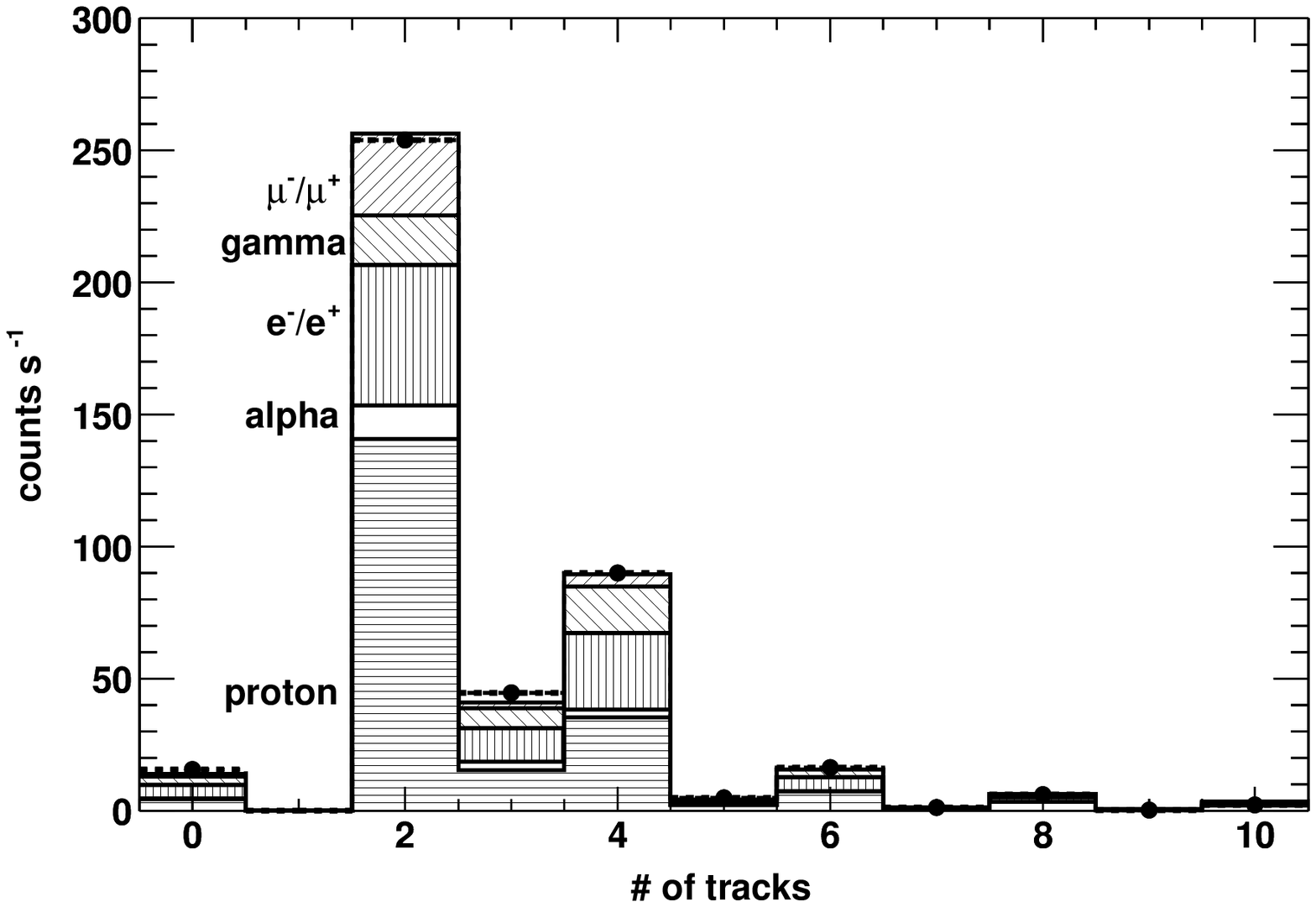}
\plotone{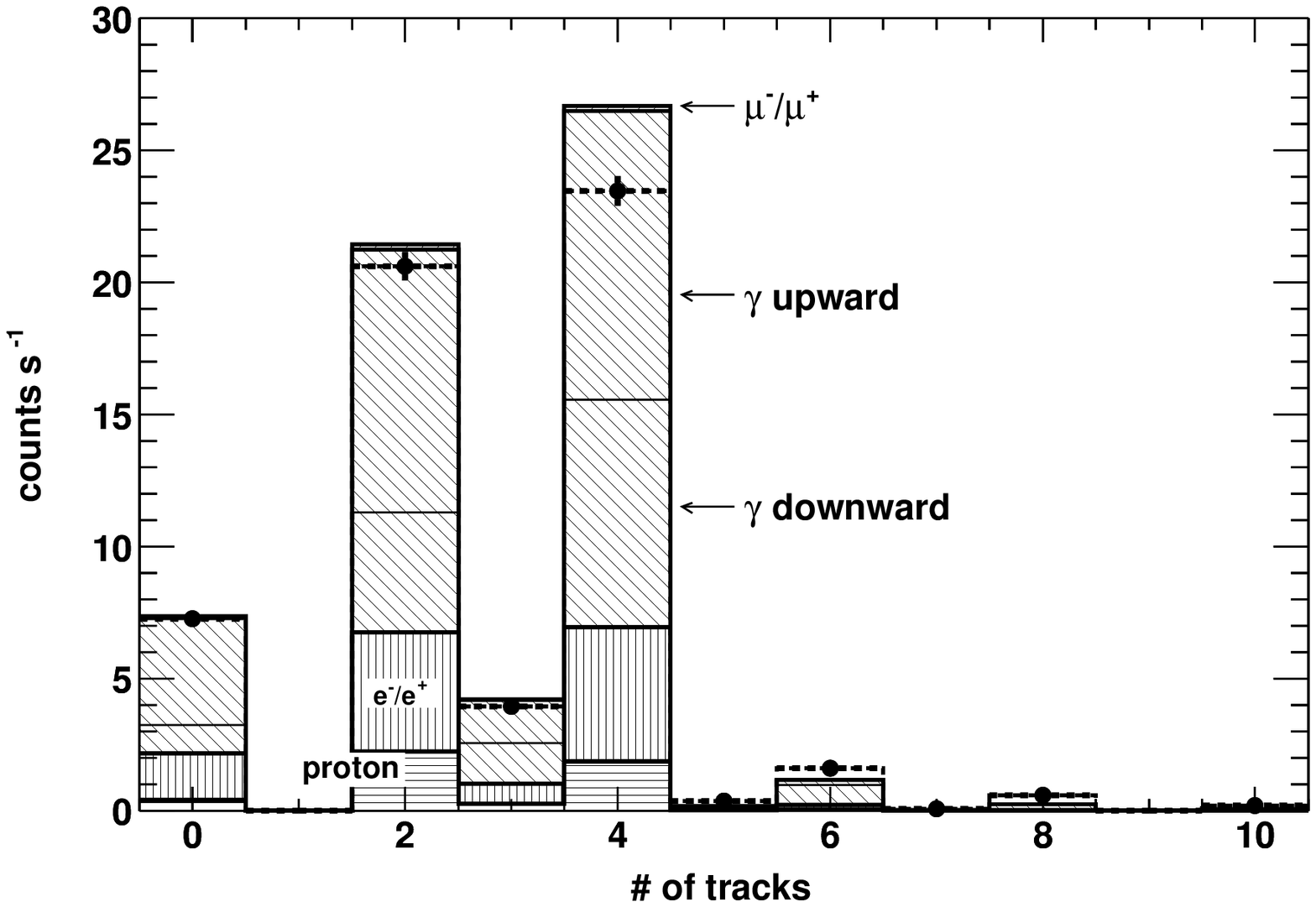} 
\caption{The number of reconstructed tracks of charged (top panel)
and neutral (bottom panel) events. 
Note that the number of tracks is two for
single track events (one in x-layer and the other in y-layer)
and four for double track events.}
\end{figure}

\begin{figure}
\epsscale{1}
\plotone{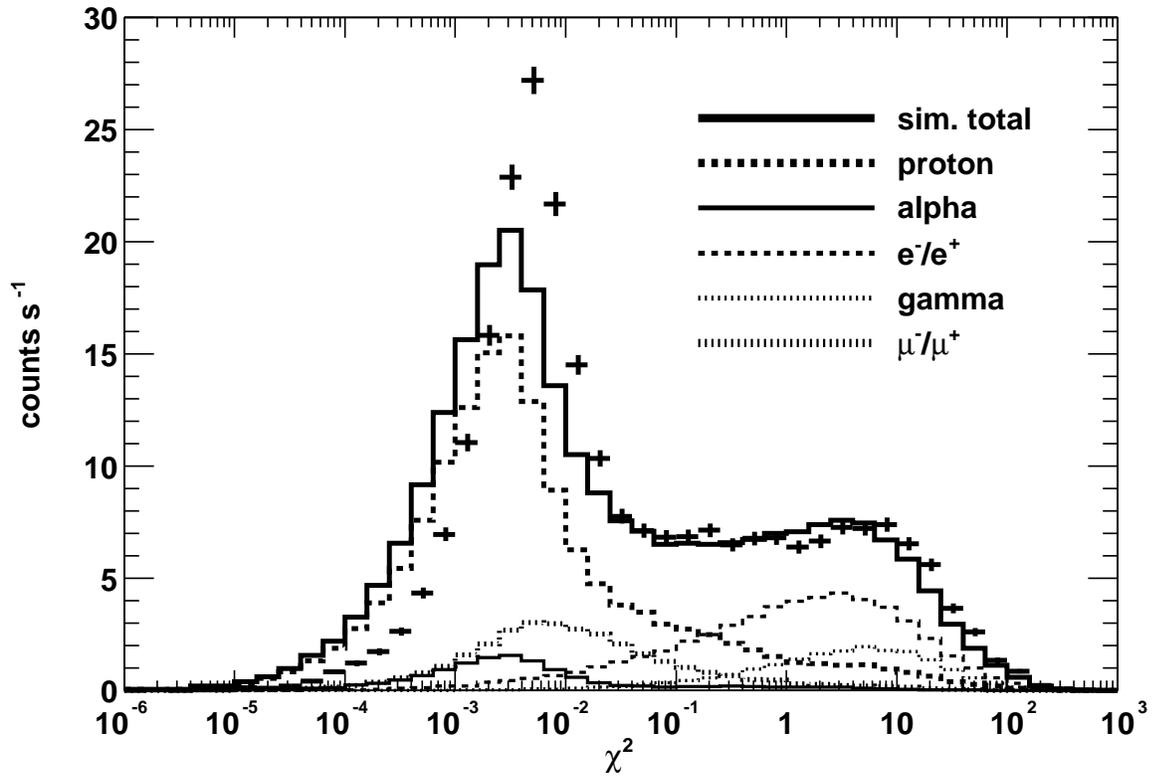}
\caption{
Comparison of the goodness-of-fit (straightness of the track)
for single track charged events between data and simulation.
}
\end{figure}

\begin{figure}
\plotone{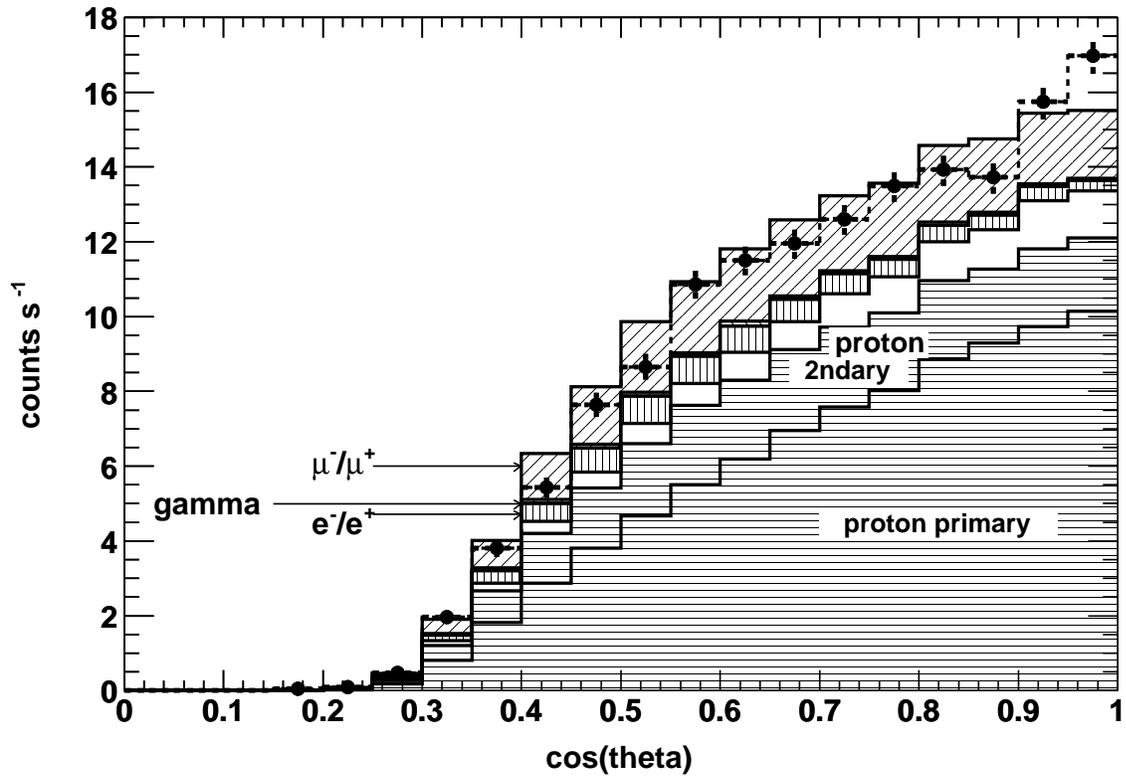}
\caption{
Zenith angle distribution of single and straight track events.
Like Figures~10 and 11, data points are given by circles and
simulation prediction by histograms. Contribution of each particle type
is also presented.
}
\end{figure}

\begin{figure}
\epsscale{0.8}
\plotone{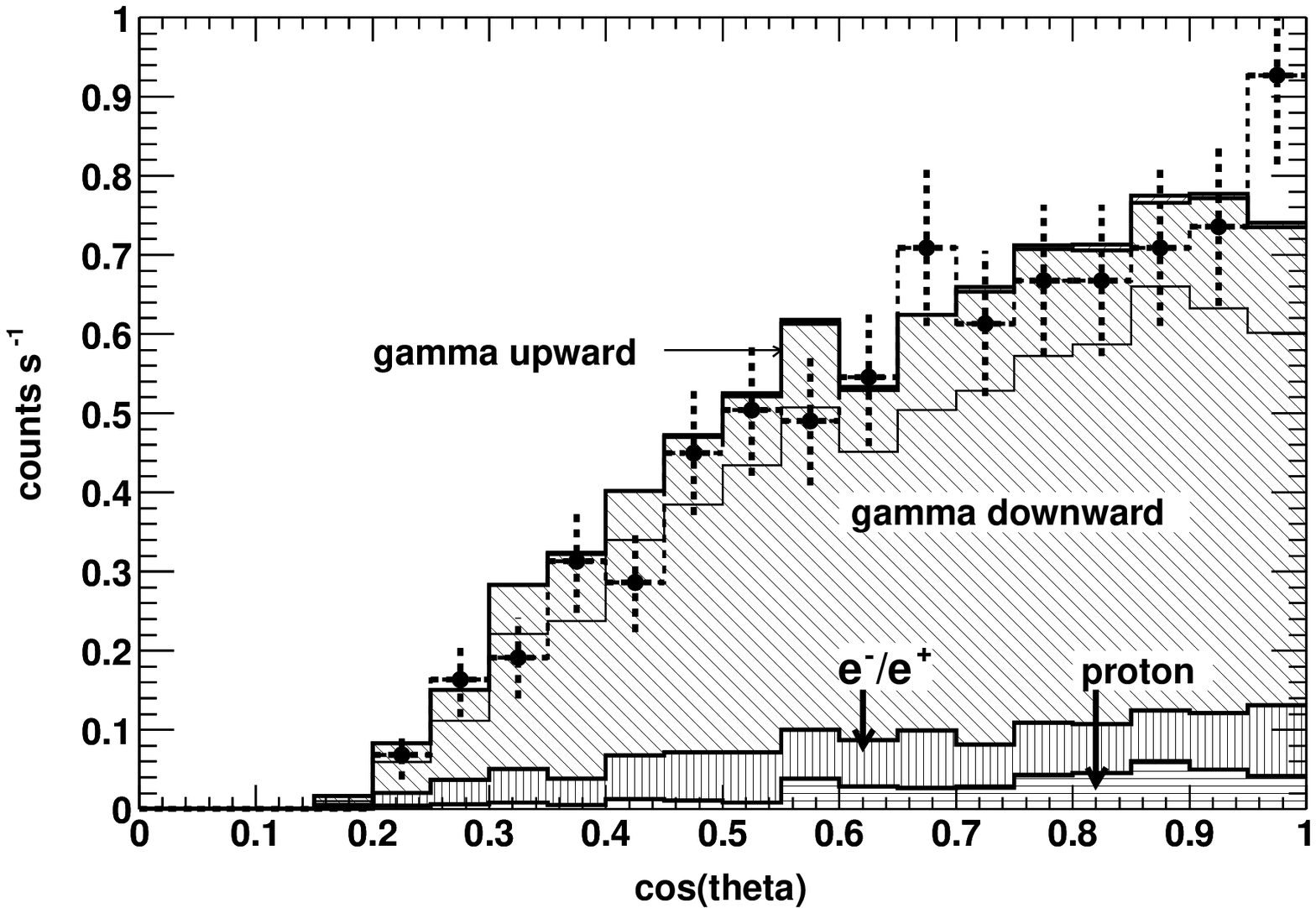}
\plotone{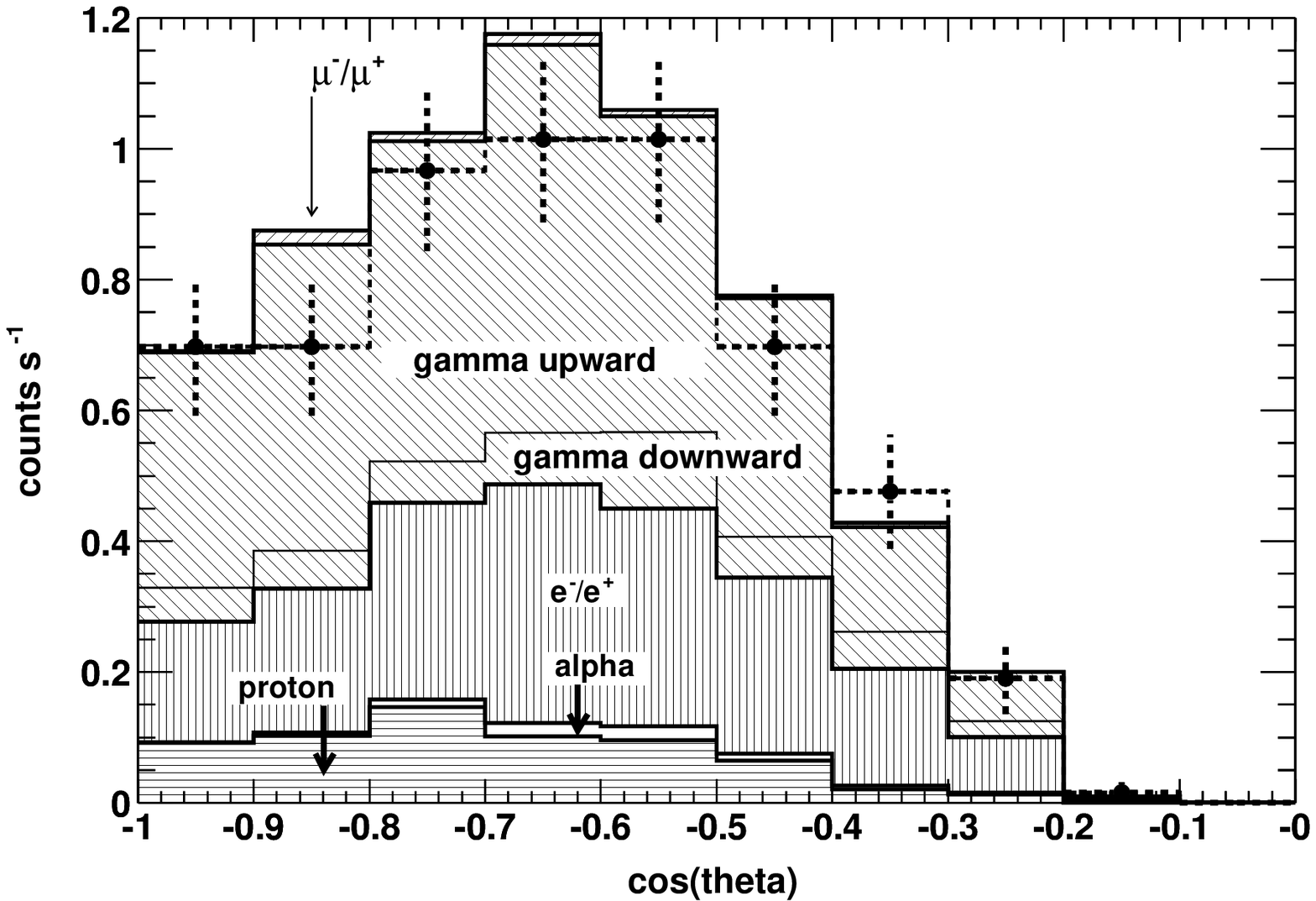}
\caption{
Zenith angle distribution of 
downward (top panel) and  upward (bottom panel) gamma-ray candidates.
Contribution of each particle type is also presented.
See text for event selection criteria.}
\end{figure}

\end{document}